\documentclass{uicthesi}
\pdfoutput=1
\input{Packages}

\begin{document}

\title{Efficient Use of Spectral Resources in Wireless Communication Using Training Data Optimization}
\author{Mohammadreza Mousaei}
\pdegrees{B.Sc., Shahid Beheshti University, Tehran, Iran, 2015}
\degree{Master of Science in Electrical and Computer Engineering}
\committee{Besma Smida, Chair and Advisor \\
Danilo Erricolo\\	
Vitali Metlushko}

\maketitle
\copyrightpage

\dedication
{\null\vfil
{\large
\begin{center}
\textit{To my mother, who believed in me when not even myself did.\\To my father, who spent his life to provide a brighter future for us.\\And to my advisor, who taught and supported me both in academia and life}\\\vspace{12pt}
\end{center}}
\vfil\null}

\acknowledgements

%There are a number of people without whom this thesis might not have been written,
%and to whom I am greatly indebted.
%\vspace{2cm}

%It is very customary to thank people who helped you finish your project/thesis. People usually include their advisers, their committee members, teammates, friends/colleagues and family members. Try to make your acknowledgments personal and specific. Here is an example from my thesis:
%
%{\color{blue}
%
I'm indebted to many people for their continuing support leading to this thesis. First, I would like to thank my adviser, Besma Smida, for her support and trust. Her supportive guidance helped me navigate through the problems and her trust gave me enough freedom to enjoy exploring the field. 
%
%I am also indebted to ....XXX and YYY..., my previous advisers at ....(your undergraduate school).... I owe them much gratitude and great respect for their guidance.
%
%Many thanks go to the Sensory Motor Performance Program (SMPP) at the Rehabilitation Institute of Chicago (RIC) for including me in their research, especially the Motor Learning and Biorobotics (MLB) meetings. 
%Furthermore, I would like to thank people in Robotics lab at the RIC for sharing their resources with my project, in particular, ...XXX... and ...YYY...
%
Being a member of Wireless lab at the University of Illinois at Chicago (UIC) was a great experience for me. The collaborations and discussions over the years helped me grow, personally and intellectually. I thank my co-authors, Besma Smida and Mojtaba Soltanalian, for their contributions in the projects. Many of the ideas in our work emerged from our discussions and teamwork. Also many thanks go to my labmates and officemates.
%
%To my parents, my siblings and my wife ....(make it personal)........
%}
%
%Remember to include your initials at the right side of the page and about 1.5 inches below the text (as follows).

\vspace{60pt}
\hfill MM

\preface
%This is an optional section. PhD students can use this part to include some legal disclaimers. For instance, you can include a copyright statement about the originality of the work done by you, the statement disclosing the funding agency that supported your work (e.g. NSF/NIH grant numbers), IRB protocol numbers for human/animal studies, previously published works and also a reference to copyright permissions that allows you to reuse those published works in your thesis. Here is an example from my thesis:
%
%{\color{blue}
This thesis is an original intellectual product of the author, M. Mousaei. All of the work presented here was conducted in the Wireless Lab at the University of Illinois at Chicago. The main project has been partially supported by National Science Foundation (NSF) under Grant Number 1620794. The results of these works have previously appeared (or is appearing) as conference publications: ICC'17 \cite{me1}, Milcom'17 \cite{me2}. The copyright permissions for reusing the published materials have been presented in Appendix. % \ref{Appendix:doppler}.

\vspace{1cm}
\begin{singlespace}
\hfill Mohammadreza Mousaei

\hfill \today
\end{singlespace}

\authorscontribution
%If you have a published work and if you have co-authors on those published works, you need to include this section. Here, for each publication that you are reusing, you need to explicitly mention the contribution of each author. Here is an example:
%
%{\color{blue}
A version of Chapter \ref{Chapter3} has been published in IEEE International Conference on Communication \cite{me1}. I was responsible for developing the majority of the ideas, mathematically representing the optimization problem,  conducting the simulations and comparing the results and finally, composing the manuscript. \\
A version of Chapter \ref{Chapter4} has been published in IEEE Military Conference on Communication \cite{me2}. I was responsible for developing the majority of the ideas, mathematically representing the optimization problem,  representation of optimization algorithm, evaluating the performance through simulations and finally, composing the manuscript.
%}

\tableofcontents
\listoftables
\listoffigures

\listofabbreviations
\begin{list}
{}
{\setlength
   {\labelwidth}{1in}
    \setlength{\leftmargin}{1.5in}
    \setlength{\labelsep}{.5in}
    \setlength{\rightmargin}{\leftmargin}}

\item[\textbf{WSN}]  \textbf{W}ireless \textbf{S}ensor \textbf{N}etworks

\item[\textbf{AWGN}]  \textbf{A}dditive \textbf{W}hite \textbf{G}aussian \textbf{N}oise

\item[\textbf{CSI}]  \textbf{C}hannel \textbf{S}tate \textbf{I}nformation

\item[\textbf{SNR}]  \textbf{S}ignal (to) \textbf{N}oise \textbf{R}atio

\item[\textbf{MMSE}]  \textbf{M}inimum \textbf{M}ean \textbf{S}quared \textbf{E}rror

\item[\textbf{DMC}]  \textbf{D}iscrete (to) \textbf{M}emmoryless \textbf{C}hannel

\item[\textbf{BEC}]  \textbf{B}inary \textbf{E}rasure \textbf{C}hannel

\item[\textbf{BSC}]  \textbf{B}inary \textbf{S}ymmetric \textbf{C}hannel

\item[\textbf{BPSK}]  \textbf{B}inary \textbf{P}hase \textbf{S}hift \textbf{K}eying

\item[\textbf{QPSK}]  \textbf{Q}uadratic \textbf{P}hase \textbf{S}hift \textbf{K}eying

\item[\textbf{CDMA}]  \textbf{C}ode \textbf{D}ivision \textbf{M}ultiple \textbf{A}ccess

\item[\textbf{PSAM}]  \textbf{P}ilot \textbf{S}ymbol \textbf{A}ssisted \textbf{M}odulation

\item[\textbf{BER}]  \textbf{B}it \textbf{E}rror \textbf{R}ate

\end{list}

\summary
Wireless communication applications has acquired a vastly increasing range over the past decade. This rapidly increasing demand implies limitations on utilizing wireless resources. One of the most important resources in wireless communication is frequency spectrum. This thesis provides different solutions towards increasing the spectral efficiency. The first solution provided in this thesis is to use a more accurate optimization metric: maximal acheivable rate (compared to traditional metric: ergodic capacity) to optimize training data size in wireless communication. Training data symbols are previously known symbols to the receiver inserted in data packets which are used by receiver to acquire channel state information (CSI). Optimizing training data size with respect to our proposed tight optimization metric, we could achieve higher rates especially for short packet and ultra reliable applications. Our second proposed solution to increase spectral efficiency is to design a multifunction communication and sensing platform utilizing a special training sequence design. We proposed a platform where two training sequences are designed, one for the base-station and the other for the user. By designing these two training sequence such that they are uncorrelated to each other, the base station will be able to distinguish between the two training sequence. Having one of the sequences especially designed for radar purposes (by designing it such that it has an impulse-like autocorrelation), the system will be able to sense the environment, transmit and receive the communication data simultaneously.  

% Chapter 1

\chapter{Introduction} % Main chapter title

\label{Chapter1} % For referencing the chapter elsewhere, use \ref{Chapter1} 

Wireless Communication has a tremendous advance in the course of recent decades. However, the improvement in the present technology requires considerably more expanding request in the sense of efficient use of resources. Such an expanding demand requires extraordinary research in this area. As wireless communication applications become more advanced and broadly utilized, the demand for more efficient use of resources is significantly increasing. One of the most important resources in wireless communication is the frequency spectrum. Since available bandwidth of frequency spectrum is limited, growing demand for wireless applications requires ideas to use spectral resources more efficiently.

\section{Existing Approaches Toward Spectral Efficiency Improvement}

In order to satisfy the demand for higher spectral efficiency, new concepts of efficient use of spectral resources   should be conceived and traditional approaches to use spectrum in wireless communication should be rethought. Here we introduce some of the existing ideas which can be utilized to achieve higher spectrum efficiency.

\subsection{Training Data Size Optimization}

%In wired communication, channel parameters (the wire) are designed, therefore the Channel State Information (CSI) is perfectly known to the receiver side. Knowing the CSI at the receiver side is crucial for receiver to be able to recover the original data. Therefore, the wired communication has a pretty good performance. On the other hand, 

% In short-packet communications, low packet efficiency is a concern: a packet typically carries less than 40\% to 50\% of actual data and the relative proportions allocated to different portions must be carefully optimized \cite{short}. 

%%%%%%%%%%%%%%%%%%%%%%
One of the primary issues in wireless communication is that Channel State Information (CSI) is obscure to the receiver. Not only the channel is unknown, but also it varies over time. This issue emerges the demand to estimate the channel in time. Channel estimation in wireless communication can be done utilizing training symbols. Training symbols are previously known symbols to the receiver. These symbols are inserted periodically into the data stream (in order to capture time variations of the channel). At the receiver side, channel can be estimated comparing the received training symbols and the known values of them. Clearly, the more training symbols we have, the better estimation of the channel we can obtain, resulting in a higher communicarion rate. However, if we increase the number of training symbols, it will cost us to loose the actual data symbols and hence decreasing data rate. Therefore, between the training data size and the communication rate, there exists a trade-off. In this manner, it is crucial to design and optimize the training symbols to achieve the optimal training data size and consequently higher rate and spectral efficiency. Traditional approaches towards this end, although efficient for general applications, are not specifically designed for particular applications (such as short packet communication). Therefore, one way to achieve more efficient use of spectral resources can be optimizing training data for specific applications.

\subsection{Tight Optimization Metric}

Wireless network research has traditionally focused on increasing the information rate to meet the demand generated by human-operated mobiles. However, all sorts of autonomous machines "things" with communication capabilities will soon need to be connected as well. The data transmitted to and from autonomous machines is very different from the data to and from human-operated mobile devices. The autonomous machines exchange a massive number of short data bursts at moderate data rates but with stringent reliability requirements. These data bursts may result from industrial automation, wireless coordination among vehicles (self driving cars), smart grid control functions, wireless sensor networks (WSN) or health-monitoring activities (wearable health-monitoring devices).  The central challenge with these new wireless services is that current wireless systems are not properly designed to support high-reliable short-packet transmission. Traditional wireless communication approaches try to optimize different parameters in wireless communication platform (such as training data, transmit power, the ratio of power allocated to training data, etc) with respect to channel ergodic capacity which leads to inaccurate results especially in short packet applications. Thus, a tighter and more accurate optimization metric is needed  for short packet communication. Since traditional wireless communication optimization approaches are not designed for short packet communication applications, they are not optimal in this area, thus, spectrum is not used optimally. Changing this metric is another way of having impact on spectrum efficiency.

\subsection{Designing Multifunction Systems}

Another traditional approach in wireless application is individual design for different wireless application. Traditionally, various applications in wireless communication (e.g. data transfer, radar and sensing applications, broadcasting, etc) had to be individually designed for specific platforms. Spectrum efficiency can be increased tremendously by designing multifunction platforms. In particular, designing a single platform for various wireless applications such as radar and communication purposes is a way of increasing spectrum efficiency. Therefore, another solution which is discussed in this thesis is to design a multifunction platform to add sensing abbilities to an already existing communication system in order to increase spectral efficiency .\\

\section{Thesis Motivations}

Some of various existing approaches towards increasing spectral efficiency has been mentioned in previous section. In this section, we will introduce our solutions to address the problem of enhancing spectral efficiency. 

\subsection{Training Data Size Optimization for Short Packet Applications}

Previously, we discussed that two major approaches to achieve higher spectral efficiency are optimizing Training Data Size and using more accurate optimization metrics especially for specific applications. Combining these two ideas together, Our first approach to increase spectral efficiency is to Optimize Training Data Size for Short Packet Applications. To that end, we optimize the training data size and investigate the dependence of this training data size on packet size and the probability of error, especially for ultra-reliable and short-packet transmission applications. This solution is explained in detail in Chapter \ref{Chapter3}. In particular, a point to point communication model where one sensor or node sends messages to a central node, or base-station,
over Rayleigh fading AWGN channel is considered. Then, the optimization problem is formalized in terms of approximate achievable rates at a given block length, training data size, and the probability of error, which results in more accurate training data size optimization. Our Simulation results in Chapter \ref{Chapter3} show that while optimizing the training data size, it is critical consider the packet size and the error probability.

%we will use tight maximum achievable rate approximations for short packet communication as our optimization metric, then try to optimize the training data size using this new metric. Our results show increased efficiency with respect to different parameters such as rate, signal to noise ratio, error probability, etc.

\subsection{Designing an Integrated Radar and Communication Platform using Training Data}

As mentioned in previous section, another way of gaining spectral efficiency is to design multifunction platforms. Such a platform to integrate sensing abilities to an already built communication system is explained in Chapter \ref{Chapter4} where we design an integrated communication and sensing platform by utilizing training data design techniques. We utilize two different
training data sequences (one is used for uplink and the other one is designed for downlink) with the
condition that they must be uncorrelated to each other. Within
such a framework, the signal received from the user and the backscattered
signal from the desired objects have uncorrelated training data sequences.
Therefore, the base-station is able to distinguish between the data sequence from user
and the back-scattered sequence from object. We assume a time division
duplex (TDD) framework. The pilot sequences are designed for
MIMO channels. We evaluate channel MSE as a figure of merit for
communication system. We also show that the designed training data sequences are
uncorrelated for a range of time lags. Moreover, designed uplink
training data sequence has negligable autocorrelation for a range of time lags leading
to an impulse-like autocorrelation for radar sensing.

\section{Notation}

In this thesis, vectors/sequences are demonstrated with lowercase bold letters and uppercase bold letters represent matrices. The Hermitian transpose, the complex conjugate, and the transpose of the vector/matrix are represented respectively with $(\cdot)^H$, $(\cdot)^*$ and $(\cdot)^T$. The $l_n$-norm of the vector $\boldsymbol{x}$  (showed as $\| \boldsymbol{x} \|_n$) is defined as $\left( \sum_k |\boldsymbol{x}(k)|^n \right)^\frac{1}{n}$ in which  $\{ \boldsymbol{x}(k) \}$ are the entries of $\boldsymbol{x}$. for a matrix $\boldsymbol{X}$, the Frobenius norm is represented as $\| \boldsymbol{X} \|_F = \left( \sum_{k,l} |\boldsymbol{X}(k,l)|^2 \right)^\frac{1}{2}$ when entries of the matrix are represented as $\{ \boldsymbol{X}(k,l) \}$.   Finally, the set of real and complex numbers are represented by $\mathbb{R}$ and $ \mathbb{C} $ respectively. %for any real number $x$, the function $[x]$ yields the closest integer to $x$ (the largest is chosen when this integer is not unique). 

\section{Thesis Structure}

The thesis can be organized as follows: Chapter \ref{Chapter2} gives a background about previous works and explains the thesis contributions compared with previous works. Chapter \ref{Chapter3} introduces our new optimization method for finite blocklength and compares its results through simulations. In Chapter \ref{Chapter4}, a novel idea for integrating communication and sensing abilities in a single platform is proposed; the idea is formulated and then investigated through simulation. Finally, Chapter \ref{Chapter5} concludes the thesis along with mentioning limitations of our research and providing solutions for those limitations as future work.

%\myPart{I}{Preliminaries}{4cm}

% Chapter 2

\chapter{Previous Works} % Main chapter title

\label{Chapter2} % For referencing the chapter elsewhere, use \ref{Chapter2} 

This research focuses on three major topics in areas of wireless communication and information theory. First, the area of optimizing training data, which has extensively gained the attention of researchers in wireless communications. Second, the area of finding asymptotically tight expression for maximum achievable rate , which dates back to the dawn of information theory itself, but has recently attracted broad research because of its applications. Finally, the area of joint radar and communication systems, that recently has gained extensive research due to the effort to increase bandwidth efficiency. This chapter tries to give a brief background about these areas of research and investigate some of previous works that has been done in these areas.

\section{Optimizing Training Data}

One of the earliest studies on the effect of training data on a channel capacity is \cite{firstTraining}, where It presents closed form expressions for the bit error rate (BER) in pilot assisted BPSK and QPSK modulations, for an upper bound on error rate in 16QAM, and for the optimized receiver coefficients. Another pioneer study of training data is \cite{thirdTraining}, where they have optimized number of training data symbols specifically for CDMA channels for both high and low rates based on channel capacity. Wireless data links promise high data rates with low probability of error when the receiver khows the Channel State Information (CSI). The receiver requires the transmitter to send the known training data during the training time (a portion of transmission time) to acquire CSI. One other factor which can contribute towards efficiency is placing of training data symbols in the packet. The optimal placing of the training data is discussed in \cite{secondTraining} where the expression in closed-form for the MMSE estimate of the channel assess as a function of training data  arrangement for block fading channels is acquired.

%\cite{hassibicite3}, where it is shown that if the number of transmit antennas is chosen so that it maximizes the throughput in the wireless channel, the length of training data should be half the coherence time.

 The concept of optimizing training data is investigated extensively in \cite{mmse3}, where the relation between training data and the capacity of fading channel is being investigated. As it is discused previously in Chapter \ref{Chapter1}, the basic concept about this relation is that there is a trade-off between the performance of the wireless system and the training data; and there is an \textit{optimal} point in which the best performance is acquired. \cite{mmse3} investigates this trade-off, and find the optimal point by optimizing over a average ergodic channel capacity. They also optimize the power allocation for training phase versus the data transmission phase then find the optimal training data that is required for the optimal power. \cite{mmse3} investigates the optimization in flat fading channels. The training symbol power is fixed in more common systems. Therefore, the optimization is over the quantity of training symbols. In that case, some explicit outcomes have been set up in both low and high power regimes. Another investigation of training data size optimization is \cite{mmse2} where a closed form answer for the average rate of training data in block fading is proposed. More recently, the optimization of the training data size in a unified continuous and block fading model is researched in \cite{lozano2008,jindal2010}. They also investigate continuous fading channels and the effects of Doppler Frequency on training data size and power allocation and the dependence of the optimum training data size on different parameters of the system (e.g. signal-to-noise ratio, fading rate) is quantified.

One of the most recent works on this topic is \cite{newHowMuch}. In this article, authors considered a more practical scenario where $K$ users are using a $M$ antenna MIMO system. Then both the number of users and the number of training data are optimized to achieve maximal ergodic capacity. Unlike conventional training-based systems, optimization of one-bit training-based system is investigated in \cite{HowMuchOne}, where one-bit analog-to-digital converters (ADCs) is used. Authors in \cite{HowMuchOne} show that optimal training length in such a system is greater than conventional ones. A brief comparison of the development of the researches in this area is also provided in Table \ref{table:Training}.

%\section{Related works}

%The optimization of training data, predicated on the maximization of the \textit{ergodic channel capacity}, has been largely studied in the literature \cite{cavers1991,medard2000,balter2001,mmse3,mmse1,mmse2,opt_tr2,ce_bem3,dong2004,furrer2007}.  
% If the SNR is high enough than the number of training symbols has to be minimal and if the SNR is sufficiently low than half the transmission packet have to be dedicated to training \cite{mmse3}.   

\begin{longtable}{|l||X|X|X|}
\caption{RELATED WORKS: OPTIMIZING TRAINING DATA.}\label{table:Training} \\

\hline \multicolumn{1}{|c||}{\textbf{ }} & \multicolumn{1}{c|}{\textbf{Citation}} & \multicolumn{1}{c|}{\textbf{Model}} & \multicolumn{1}{c|}{\textbf{Results}} \\ \hline 
\endfirsthead

\multicolumn{4}{c}%
{\tablename\ \thetable{} -- continued from previous page} \\
\hline \multicolumn{1}{|c||}{\textbf{ }} & \multicolumn{1}{c|}{\textbf{Citation}} & \multicolumn{1}{c|}{\textbf{Model}} & \multicolumn{1}{c|}{\textbf{Contribution}} \\ \hline
\endhead

\hline \multicolumn{4}{|r|}{{Continued on next page}} \\ \hline
\endfoot

\hline
\endlastfoot

 & \cite{firstTraining} & Gaussian Fading Channel & Closed form expressions for the bit error rate (BER) in pilot assisted BPSK and QPSK \\
Early Works & \cite{thirdTraining} & CDMA with AWGN Multi-path Fading Channel & Optimizing number of training data symbols specifically for CDMA \\
 & \cite{secondTraining} & Gaussian Fading Channel  & Closed form expressions for the bit error rate (BER) in BPSK and QPSK \\
 \hline
 & \cite{mmse3} & Rayleigh block fading channel & Optimizing number of training data symbols and power allocation \\
Recent Works & \cite{mmse2} & PSAM with Different Fading Models & Closed form MMSE for Fading Channels  \\
 & \cite{jindal2010} & PSAM with Block and Continuous Fading & Closed form for optimal number of training data symbols for both continuous and  \\
 \hline
 Most Recent & \cite{newHowMuch} & K users which are using a Rayleigh Fading MIMO system & Finding closed form expression for optimal number of training data symbols and optimal number of users  \\
 Most Recent & \cite{HowMuchOne} & One bit modulation with Rayleigh Fading & Finding closed form expression for optimal number of training data with one bit modulation \\

\end{longtable}

\section{Finite Blocklength Regime}

The problem of precisely computing the rate for various communication channels has the most measure of research from the dawn of communication theory. In this section, the earliest works in this area will be introduced and their contributions will be discussed. Then recent works towards finding a tight approximation for maximum achievable rate will be introduced. Finally, the most recent works in this area will be discussed.

%the fundamental properties of channel will be introduced, then a brief description of the capacity and rate for a simple AWGN Discrete Memoryless Channel (DMC) is given. Finally, more extensively, the same principles will be discussed about fading channels, which is the channel model for this thesis

Without doubt, the most fundamental property of a communication channel is the capacity introduced by Shannon \cite{shannon}. The maximal bound for achievable rate with a vanishing probability of error when the codeword length goes to infinity is called the Channel Capacity. This fundamental bound was introduced by Shannon for different channels, for example, the additive white gaussian noise channel (AWGN) an the discrete memoryless channel (DMC) \cite{shannon}. The basic problem with Shannon capacity is that it cannot express the relation between a target error probability and the corresponding blocklength, i.e, it doesn't answer the question of "how much large should the blocklength be to attain a certain probability of error?" and it simply states that if the blocklength is "large enough", the probability of error vanishes to zero.

%This section will shortly describe the idea behind the channel capacity, at that point, elaborate more about another fundamental property of the channel named channel dispersion. The goal of introducing channel dispersion is to approximate the communication rate more precisely. Therefore, the communication rate will be introduced in this section and its relation with channel dispersion will be discussed.

The behavior of the communication rate with respect to other parameters of interest, i.e. achievable rate for limited blocklength, is an essential problem going back to to Feinstein and Gallager in the 60's \cite{feinstein,galla}. Feinstein's acheivibility lower bound on rate is introduced in \cite{feinstein} and Gallager's random coding lower bound on average probability of error is introduced in \cite{galla}. Recently, Polyanskiy \textit{et al.} reformulated the issue and altogether determined some non-asymptotic achievability and converse bounds for the finite blocklength regime \cite{poly}. They showed that for various channels, the data rate
varies with packet sizes, desired error probabilities, and
channel dispersions. The channel dispersion for various channel models is investigated by Polyanskiy. In particular, the channel dispersion for coherent fading channels is investigated in \cite{polydmc} and closed form for channel dispersion is derived. 

The most recent researches on this area includes \cite{newFinite1,newFinite2}. Where finite blocklength bounds on the maximum achievable rate when channel state information is obscure both from transmitter and receiver in MIMO Rayleigh block-fading channels is presented in \cite{newFinite1} and a closed form expression for channel dispersion in MIMO block-fading channels is derived in \cite{newFinite2}. A brief comparison of the development of the researches in this area is also provided in Table \ref{table:Metric}.

\begin{longtable}{|l||X|X|X|}
	
\caption{RELATED WORKS: FINITE BLOCKLENGTH REGIME.} \label{table:Metric} \\

\hline \multicolumn{1}{|c||}{\textbf{ }} & \multicolumn{1}{c|}{\textbf{Citation}} & \multicolumn{1}{c|}{\textbf{Model}} & \multicolumn{1}{c|}{\textbf{Contribution}} \\ \hline 
\endfirsthead

\multicolumn{3}{c}%
{\tablename\ \thetable{} -- continued from previous page} \\
\hline \multicolumn{1}{|c||}{\textbf{ }} & \multicolumn{1}{c|}{\textbf{Citation}} & \multicolumn{1}{c|}{\textbf{Model}} & \multicolumn{1}{c|}{\textbf{Contribution}} \\ \hline
\endhead

\hline \multicolumn{4}{|r|}{{Continued on next page}} \\ \hline
\endfoot

\hline
\endlastfoot

 & \cite{shannon} & DMC channels and Discrete channels with finite memory & Showing that for a channel with vanishing error and infinite blockingth, rate has an upper bound $C$ \\
Early Works & \cite{feinstein} & General Coding Theory Bound & Introducing an achievability lower bound leading to a relationship between rate and probability of error \\
 &\cite{galla} & Coding Theory Bound applied on BSC and AWGN channels & Introducing a random coding lower bound leading to a relationship between probability of error and channel coding length \\
 \hline
Recent Works & \cite{poly} & Binary Erasure Channel (BEC), BSC and AWGN channels & Finding tight approximation of maximum achievable rate in terms of channel capacity and channel dispersion \\
 & \cite{polydmc} &  SISO coherent fading AWGN channel & Finding closed form expression for channel dispersion in fading channels  \\
 \hline
 Most Recent & \cite{newFinite1} & MIMO Rayleigh block-fading channel
  & Presenting finite blocklength bounds on the maximum achievable rate in absence of CSI from both transmitter and receiver. \\
 Most Recent & \cite{newFinite2} & MIMO block-fading AWGN channel & Finding closed form expression for channel dispersion in MIMO block-fading channels.\\

\end{longtable}

\section{Integrated Radar and Communication}

Due to the increasing demand in wireless communication
services, achieving higher data rates and more reliable trans-
missions have become a fundamental goal . Considering the consistently expanding interest for both high-speed communication services and accurate remote sensing capabilities, present day wireless frameworks will progressively require more effective procedures for utilization of the accessible frequency spectrum. In particular, the coexistence of communication and radar systems has recently attracted a significant research interest. Using the same frequency spectrum for both communication applications and radar tasks can  significantly increase spectral efficiency. That is why the integration of various functions such as communication applications with navigation and radar tasks has attracted substantial interest in recent years.

The idea of combining communication and radar systems in a single platform has been proposed in \cite{idea} and such a system has been developed and implemented in \cite{implement}. One of early works in integration of radar and communication platforms is the work by Office of Naval Research that propelled the Advanced Multifunction Radio Frequency Concept (AMRFC) program \cite{AMRFC}. The AMRFC program was persuaded by the absence of integrated radar, communication platform. The idea proposed in this work went for reasonable broadband RF gaps that can deal with concurrent operation of different functionalities and is focused on the RF front-end. Thereafter,  extensive research has been done to this end. Exploiting the main lobe of the beam for radar purposes, and the sidelobes (which are of no significance to the radar pulse compression) for data transmission purposes is investigated in \cite{sidelobe}. \cite{intrapulse,timemod} approach the same problem by devising similar methods to allow comparably low data rates into an already existing radar system. Assuming for Global Navigation Satellite System (GNSS) \cite{GNSS} utilizes weighted pulse trains for data signal and analytically shows that in such system, data transfer can be integrated. 
Using OFDM signal structure to construct a joint radar-communication system (RadCom) is discussed in the literature \cite{OFDM2}, mainly by utilizing the distortion of back-scattered OFDM signal compared to transmitted OFDM signal to achieve range and Doppler profiles of the object. OFDM based RadCom is extended for multipath and multiuser scenario in \cite{OFDMRadComExt}. A MIMO OFDM RadCom system is proposed in \cite{RadComMIMO}, in which a OFDM waveform is designed to be suitable for performing both radar sensing and data transmission.  Contrary to former literatures, RadCom has higher data rates. A brief comparison of the development of the researches in this area is also provided in Table \ref{table:RadCom}.

\begin{longtable}{|l||X|X|X|}
	
\caption{RELATED WORKS: INTEGRATED RADAR AND COMMUNICATION.}\label{table:RadCom} \\

\hline \multicolumn{1}{|c||}{\textbf{ }} & \multicolumn{1}{c|}{\textbf{Citation}} & \multicolumn{1}{c|}{\textbf{Technique}} & \multicolumn{1}{c|}{\textbf{Contributions}} \\ \hline 
\endfirsthead

\multicolumn{4}{c}%
{\tablename\ \thetable{} -- continued from previous page} \\
\hline \multicolumn{1}{|c||}{\textbf{ }} & \multicolumn{1}{c|}{\textbf{Citation}} & \multicolumn{1}{c|}{\textbf{Technique}} & \multicolumn{1}{c|}{\textbf{Contributions}} \\ \hline 
\endhead

\hline \multicolumn{4}{|r|}{{Continued on next page}} \\ \hline
\endfoot

\hline
\endlastfoot

 & \cite{idea} & Spread Spectrum Technique & Proposing a simulation prototype of a vehicle to vehicle communication and ranging system \\
Early Works & \cite{AMRFC} &  Utilizing
a set of broad-band array antennas & Introduced suitable broadband RF apertures for integrating communication and radar platforms \\
 & \cite{GNSS} & Utilizing weighted pulse trains  & Analatical analysis of using weighted pulse trains for communication and radar integration \\
 \hline
 & \cite{OFDM2} & Utilizing the distortion of backscattered OFDM signals  & Acheiving range and doppler profiles of object in a communication system \\
Recent Works & \cite{mmse2} & PSAM with Different Fading Models & Closed form MMSE for Fading Channels  \\
 & \cite{RadComMIMO} & Special OFDM waveform design & OFDM waveforms are designed specifically for integration of radar and communication system  \\
 \hline
 Most Recent & \cite{timemod} & Utilizing time modulated arrays &  Specifying mainlobe for radar and sidelobes for communication purposes using time modulated arrays \\
 Most Recent & \cite{sidelobe} & Utilizing antenna lobes & Exploiting the main lobe of the antenna for radar purposes and the side lobes for communication purposes  \\

\end{longtable}

\section{Contribution}

In this section, we briefly explain the departure of our work from previous ones and how the distinction of our work can help improving the spectral efficiency. As mentioned before, our work can be divided in two major contributions:

\subsection{Optimizing Training Data in Finite Blocklength Regime}

Optimization posed in prior works is predicated on the maximization of the \textit{ergodic channel capacity}.  They all assume that the packet error probability can be reduced by picking the packet length adequately large. This optimization, based on large block-length, is unsuitable for short-packet transmission. In short-packet communications, low packet efficiency is a concern: the relative proportions allocated to different portions (training and data) must be carefully optimized \cite{short}. Indeed, we need a new analysis of the achievable rate to assess the performance of short-packet communication \cite{poly}. Unfortunately, even for channel models that are much simpler to analyze than the one encountered in this thesis, the exact value of achievable rate is unknown \cite{short}. Polyanskiy and \textit{al.} recently provided a unified approach to obtain tight asymptotic on achievable rate by providing lower bound that coincides with an upper bound in \cite{poly}. They showed that for various channels, the data-rate varies with packet sizes, desired error probabilities, and channel dispersions. 

The key departure from prior works on pilot overhead optimization is that we (a) use asymptotically tight expression on maximum achievable-rate as the optimization metric to optimize the training data size. Therefore, our optimal training results are more accurate for short-packet transmission, (b) expressed the minimum mean square error for continuous fading in a closed form expression as function of the packet size. These results are later used to optimize the training data for continuous fading channel models, (c) investigate the dependence of the training data size on different system parameters, e.g. packet size, error probability, signal-to-noise ratio (SNR) and fading rate. Our investigations shows that our optimal training data has better performance especially in short packet communication and ultra reliable applications.

\subsection{Designing an Integrated Communication and Sensing Platform Using Training Data}

While integrating radar and communication operation in one system has been considered in the mentioned litriture in previous section, such efforts are typically centered around incorporating communication as a secondary operation alongside a primary radar operation. Some litriture like \cite{GSM} proposed a system for adding radar tasks to an already existing communication system. However, such systems are proposed for a specific communication model (\cite{GSM} is designed for GSM Communication systems). In this work, we  propose an integrated system of communication and sensing (which we call \emph{ComSens}) that relies on  the communication pilot overhead--- thus paving the way for pilot design and exploiting pilot diversity to achieve a satsifactory performance in both communication and radar tasks. As opposed to prior works, we propose a novel general system, which is not designed for a specific communication system. Our work has two primary advantage in compare with prior works: \\

\begin{itemize}

\item The first novelty of our work is that we proposed the idea of utilizing training data to achieve the integrated system. Training based channel estimation is very common. As it is mention in previous sections, accurate knowledge of channel state information (CSI) is important for wireless communication systems. Most modern wireless systems acquire the CSI with the assistance of training data. For that reason, our proposed system can be implemented in many commonly used wireless models.

\item Another distinction of our work is that we propose a system where a secondary sensing system is added to an already existing primary communication system. The reason behind this idea is that communication devices are more ubiquitous than radar systems. We note that incorporating the communication signals in the primary radar probing waveforms may not be an efficient fusion of communication and radar systems. In fact, the communication task must play a primary rule not only because of the pervasive usage of comuunication devices, but also the fact that the communication systems typically require a larger capacity of conveying information than radar systems. Additionally, considering the communication operation as the primary lays the ground for making the radar systems ubiquitous (for example having radar capability on cellphones).
 
\end{itemize}

\section{Conclusion}

This chapter discussed previous works in areas of focus in this research. In each area, earliest works has been introduced and briefly explained and discussed. Then core literature has been discussed more deeply. Afterwards, the most recent literature has been explained and finally a brief history of development in each area is provided in tables \ref{table:Training}, \ref{table:Metric} and \ref{table:RadCom}. Moreover, the contribution of our work compared with the discussed literature has been explained. It was discussed that our proposed solutions for increasing the spectral efficiency are to (a) use tight maximal achievable rate approximations as a metric to optimize training data size to be able to design more accurate systems for short packet communication systems, and (b) design a multifunction communication and sensing system, enabling the system to use frequency spectrum more efficiently. This chapter, along with chapter 1, has built a road-map to this research. In next two chapters, the details for these two contributions will be presented.

% Chapter 3

\chapter{Training Data Size Optimization for Finite-Blocklength Regime} % Main chapter title

\label{Chapter3} % For referencing the chapter elsewhere, use \ref{Chapter3} 

As mentioned before, one solution that we present through this thesis is to optimize the training data size with maximal achievable rate approximation as the optimization metric. This chapter specifically explains our method of optimizing training data size, provides optimization problem of finding optimal training data sequence size, and compares our proposed optimal training data size performance to previous methods through soultions.

\section{Preliminaries}

This section is dedicated to explain pre-requisites for explaining mathematical representation of the problem. The communication model for which we introduce our optimization method will be introduced in this section.

\subsection{Channel Model}

\begin{figure}[H]
	\centering
	\includegraphics[width = 12cm]{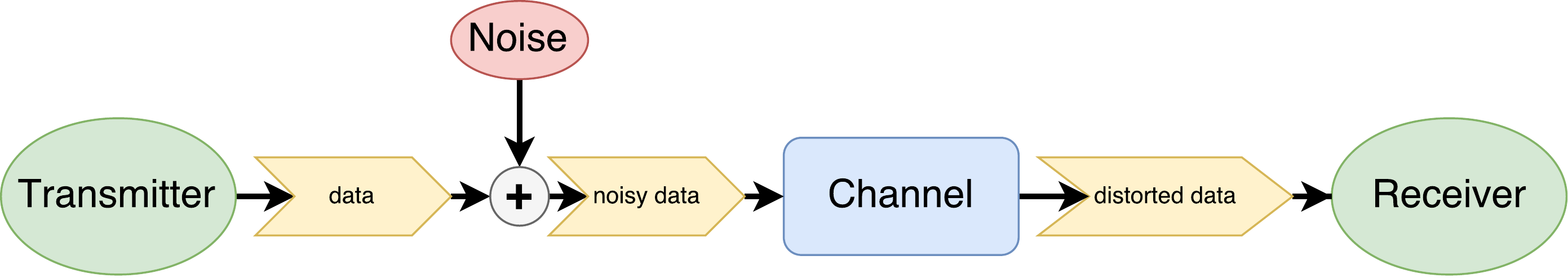}
	
	\caption{Fundamental elements in a communication framework.}
	\label{fig:Framework}
\end{figure}

In general, a communication framework comprises of four essential components, the transmitter, noise, channel and the receiver. As it is shown in Fig~\ref{fig:Framework}, the transmitter sends the data and the superposition of noise and the data goes through channel to the receiver. Data winds up distorted in the process of going through the channel. That is the reason the data should be recovered at the receiver side. Recovery occurs at the receiver side in equalizer. Equalizer is basically the inverse of the channel, so that the original data is recovered after going through equalizer. \\

In this thesis, we consider an AWGN channel with Rayleigh fading channel.  Training data symbols are periodically inserted in every packet. Assuming that each communication packet contains $n$ symbols. 
Under this model,  the relationship between input and output of $i^{th}$ received symbol can be represented as:
\begin{equation} \label{eq:chmodel} y(i) = \sqrt{\rho}x(i)h(i) + w(i), \; i = 0,1,\ldots,n-1\end{equation}
where $x(i)$ is the $i^{th}$ symbol, $y(i)$ is the corresponding received symbol, $w(i)$  is  AWGN with zero-mean. We normalize the Rayleigh fading channel ($|\mathbf{h}|^2=1$), and we assume $x(i)$ and $w(i)$ have unit mean square. Thus, $\rho$ is the SNR at the receiver.

%
%\begin{eqnarray}\label{eq:chmodel}
%y(i) &=& \sqrt{\rho}x(i)h(i) + w(i), \quad i = 1, 2, \dots, N 
%\end{eqnarray}
%where $x(i)$ is the $i^{th}$ transmitted symbol; $y(i)$ is the corresponding received symbol; $w(i)$ is the AWGN noise with zero-mean. Without loss of generality, $x(i)$ and $w(i)$ are assumed to have unit mean. Thus, $\rho$ is the Signal to Noise Ratio (SNR) at the receiver. 

\subsection{Coding Theory Model}

Coding theory has a slightly different and more mathematical definition of a communication framework. Consider a source , modeled as a random variable which equi-probably takes values in the set $\{1, \dots, M\}$. The channel is a noisy medium which takes a symbol in the alphabet $\mathcal{A}$ and output a symbol in the alphabet $\mathcal{B}$. An encoder is a mapping from messages ($\{1, \dots, M\}$) into sequences where $n$ symbols are used $\mathcal{A}^n$("codewords"). Therefore, the encoder can be considered as a function $f: \{1, \dots, M\} \rightarrow \mathcal{A}^n$. Estimation of original message by looking at sequences of channel output with length $n$ can be considered as a function $g: \mathcal{B}^n \rightarrow \{1, \dots, M\}$, which is done by the decoder. The principal goal in communication is to find an encoder-decoder pair (code) which is able to communicate messages with some fixed probability of error $\epsilon$. Such code is called ($n, M, \epsilon$)-code.

\begin{figure}[H]
	\centering
	\includegraphics[width = 6cm]{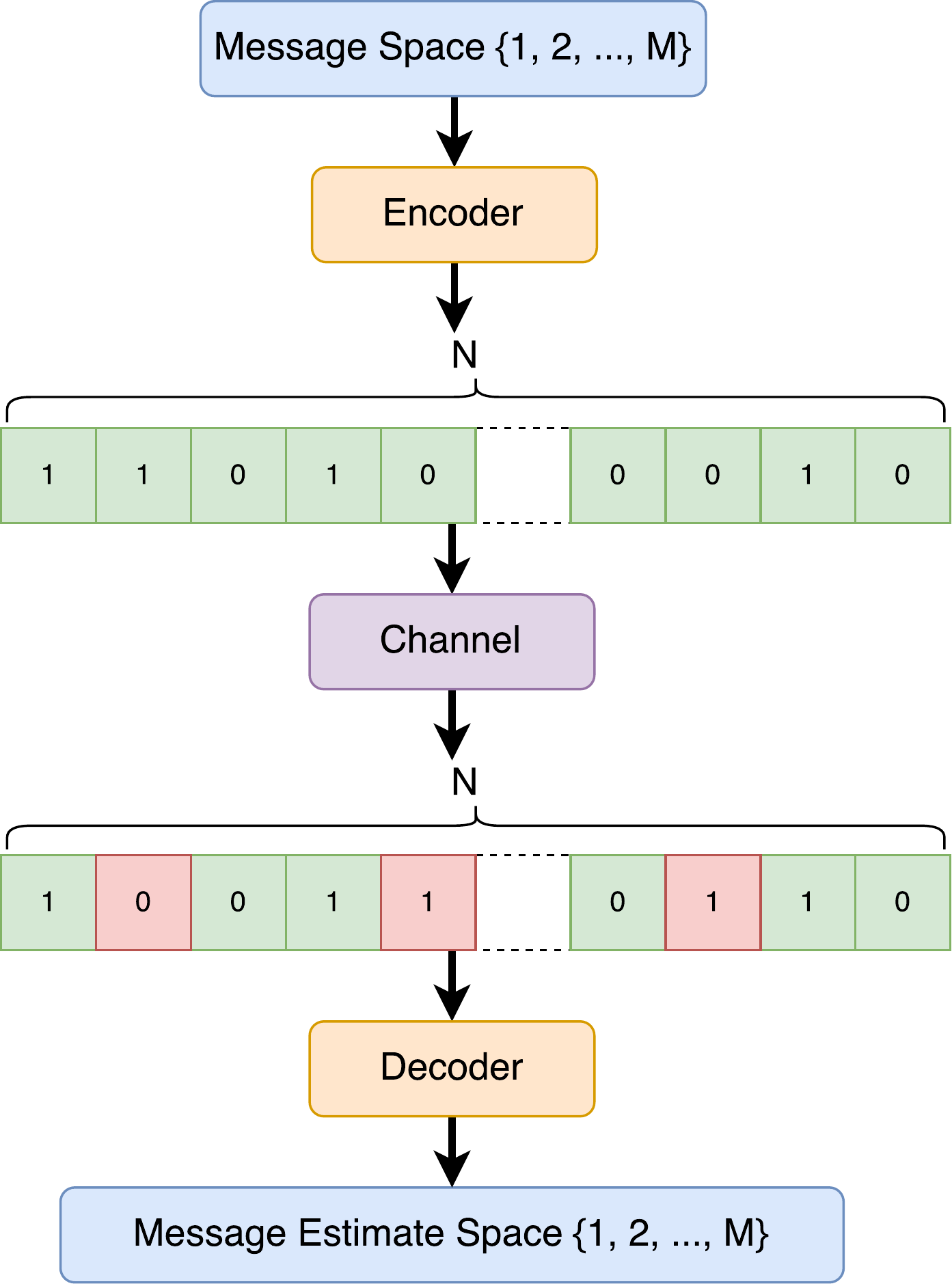}
	
	\caption{The coding theory view of communication channel.}
	\label{fig:Packet}
\end{figure}

\subsection{Packet Structure}

Information is transmitted in chunks of data called packet. As it is clarified beforehand, Packet is comprised of training symbols and data symbols. Through this thesis, we assume that the packet is of size $n$ and there are $n_t$ pilot symbols in a packet (each of which periodically inserted in the packet with period $\frac{1}{\alpha}$). Therefore, the equality $n = \frac{1}{\alpha} n_t$ holds for any $n$.

\begin{figure}[H]
	\centering
	\includegraphics[width = 12cm]{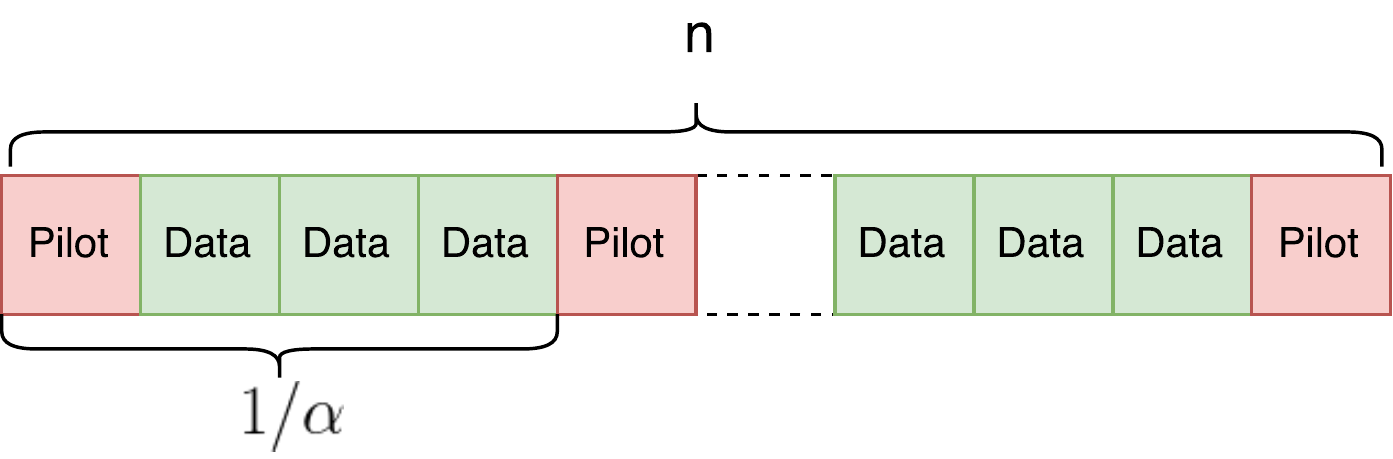}
	
	\caption{The structure of the packet.}
	\label{fig:Packet}
\end{figure}

\section{Training Data Size Optimization}

%%%%%%%%%%%%%%%% HASSINBI BEGIN %%%%%%%%%%%%%%%%%%%%%%%%%%%%

We consider a  point-to-point communication, in which  one sensor wishes to send messages to a central node, or base-station. The sensor enables the base-station to estimate the channel gain by sending training symbols known to the base-station. As it is previously mentioned in Chapter \ref{Chapter2}, %REPHRAZE THIS PART
in this section we use maximal achievable-rate tight approximation as the optimization metric which is more accurate for short-packet transmission. Then we derive the minimum mean square error for continuous fading as function of the packet size; and investigate the dependence of the training data on different system parameters, e.g. packet size, error probability, signal-to-noise ratio (SNR) and fading rate.

%\section{Training-Based Channel Estimation}

To acquire the knowledge of CSI at the receiver, transmitter sends training symbols ($x_t(i)$) to the receiver. The training symbols are already known at the receiver. Having the knowledge of transmitted training symbols $x_t(i)$ and the corresponding received symbols at the receiver $y_t(i)$, the receiver tries to acquire an estimate of the channel $\widehat{h}$. This channel estimation then is used during the data transmission as the known channel response (CSI).

\subsection{Training-Based Scheme}

The transmission is divided into 
two phases. The training phase includes $n_t$ symbols and the data transmission phase includes $n-n_t$ symbols. We define parameter $\alpha = n_t/n$\footnote{ $\alpha$ should be not less than $\alpha_{min} = 1/n$.}.

	\subsubsection{Training Phase} During this phase, training symbols are transmitted. Although the transmission of training symbols is not sequential, it is considered as a separate phase for mathematical simplicity. During this phase, we can rewrite Eq. (\ref{eq:chmodel}) as follows
	
	\begin{equation}
	y_t(i) = \sqrt{\rho}x_t(i)h(i) + w_t(i), \quad i = K, 2K, \dots, MK=N 
	\end{equation}
	where $x_t(i)$ and $y_t(i)$ are the $i^{th}$ transmitted training symbol and the corresponding received symbol respectively. The receiver in this phase utilizes $y_t$ and $x_t$ to generate the estimation $\widehat{h}$ of channel $h$.
	\\
	
	\subsubsection{Data Transmission Phase} The data symbols are transmitted during this phase. Here we may write
	\begin{equation}\label{chModel2}
	y_d(i) = \sqrt{\rho}x_d(i)h(i) + w_d(i), \quad  1 \leq i \leq N,\quad  i \neq mK\quad \forall m \in \mathsf{N} 
	\end{equation}
	where $x_d(i)$ and $y_d(i)$ are the $i^{th}$ transmitted data symbol and the corresponding received symbol respectively. The receiver in this phase tries to recover the data $x_d(i)$ using the estimated channel $\widehat{h}$ obtained in the training phase\\

\subsection{Pilot-Based Estimation}

The $n_t$ training symbols are utilized for channel $h(i)$ estimation for all $i$ in the data transmission stage. We initially assess the minimum channel estimation error of the channel vector $\mathbf{h}:=[h(0), \ldots, h(n)]^T$, as a function of $n_t$, which is needed to consequently infer the inexact achievable rate. Let $\widetilde{\mathbf{h}}=\mathbf{h}-\widehat{\mathbf{h}}$ mean the mismatch between the original channel vector $\mathbf{h}$ and the estimation of the channel $\widehat{\mathbf{h}} :=[\widehat{h}(0), \ldots, \widehat{h}(n)]^T$.
The process of acquiring an estimate $\widehat{h}$ using $x_t$ and $y_t$ differs according to channel fading model. In this work, two distinctive fading models are considered which are introduced in the accompanying.

\subsubsection{Channel Estimation with Block Fading}

The block fading model applies to a channel in which every communication packet is influenced by a similar fading value and the fading in various packets are independent and identically distributed. For instance, this model is pertinent to wearable health sensors which are transferring short-packet communication (such as body temperature and heart bit rate) to smartphones. Communication environment in such framework changes in a low speed in such a way that channel gain fluctuates so gradually with time that it can be accepted as constant along a packet. Considering this fading model, to to acquire the channel estimate $\widehat{h}$, we can use Minimum Mean Squared Error (MMSE) estimator. The MMSE channel estimate can be expressed as bellow \cite{mmse3}

\begin{eqnarray}
\widehat{\textbf{h}} &=& \sqrt{\rho} (1 + \rho|\textbf{x}_t|^2)^{-1} \textbf{x}^*_t \textbf{y}_t \nonumber\\
&=& \dfrac{1}{\sqrt{\rho}} \left( \dfrac{1}{\rho} + |\textbf{x}_t|^2 \right)^{-1} \textbf{x}^*_t \textbf{y}_t
\end{eqnarray}

where  $\textbf{x}_{t}$, $\textbf{y}_{t}$ are input and output training symbol vectors. Defining $\widetilde{\textbf{h}}$ as the channel estimation mismatch ($\widetilde{\textbf{h}} = \textbf{h} - \widehat{\textbf{h}}$), the variance of channel estimate mismatch can be expressed as below \cite{mmse3}

\begin{eqnarray}
\sigma^2_{\widetilde{h}} = \dfrac{1}{1+\alpha n \rho}
\end{eqnarray}

\subsubsection{Channel Estimation with Continuous Fading }

For wireless communication the continuous fading channel model is more realistic. Indeed, the channel is constantly changing, so the real channel will deviate progressively from the estimation of the channel acquired at the training time. The channel estimation error for the continuous fading channels occurs because of the noise and in addition by the temporal variation of the channel \cite{contfad}. We can hence model our channel as a block fading channel with an additional noise due to the temporal variation of the channel. Assuming MMSE estimator and continuous fading model, we can obtain $\sigma^2_{\mathbf{\widetilde{h}}}$ as following 
\begin{eqnarray}
\sigma^2_{\mathbf{\widetilde{h}}} = \dfrac{1}{1 + \alpha n \rho} + \sigma^2_{Doppler},
\end{eqnarray}
where  the additional channel estimation error  $\sigma^2_{Doppler}$ is derived, in the appendix, for Rayleigh fading as 
\begin{eqnarray}\label{dopplerSigma}
\sigma^2_{Doppler} &=& 2\left(\dfrac{\pi\alpha n\rho f_D}{1 + \alpha n \rho}\right)^2\left( n - \dfrac{\alpha n}{2} \right)^2.
\end{eqnarray}
where $f_D$ is the Doppler frequency normalized to the data rate. (see appendix for proof)

\subsection{Pilot-Based Detection}

%{\color{red} CAN BE SECTIONIZED}

During the last section, it was shown how a channel estimation ($\widehat{h}$) is acquired using pilot symbols during the training phase. Now that a channel realization is known to the receiver, the receiver should actually use this information in the data transmission phase to detect the data symbols. This section tries to elaborate the process of detection during data transmission phase and investigate how it affects the capacity and performance of the system.
During the data transmission phase, the channel model (\ref{chModel2}) can be rewritten as

\begin{equation}
\label{chMMSE}
y_d(i) = \sqrt{\rho}x_d(i)\widehat{h}(i) + \sqrt{\rho}x_d(i)\widetilde{h}(i) + w_d(i),
\end{equation}
where $\widetilde{h} = h - \widehat{h}$ is the channel estimation mismatch. The receiver is provided perfectly with CSI ($\widehat{h}(i)$). Eq. (\ref{chMMSE}) can be viewed as a data transmission through a perfectly known channel ($\widehat{h}(i)$), with additional noise $\sqrt{\rho}x_d(i)\widetilde{h}(i) + w(i)$. To investigate how it affects the performance, we should investigate the effects on the capacity.

%The main problem here is the fact that the noise  $\sqrt{\rho}x(i)\widetilde{h}(i) + w(i)$  includes the channel estimation error. So the noise is not necessarily independent from the transmitted signal nor Gaussian.  First, we assume MMSE estimator, then $\widetilde{h}(i)$ and $\widehat{h}(i)$ are orthogonal. Then we consider Gaussian noise, following the same approach used in ~\cite{mmse3}. Using those assumptions,  the channel defined in Eq. (\ref{chMMSE}) became similar to the channel introduced in Section II-C, with SNR 
%\begin{eqnarray}
%\rho_{e} &=&  \dfrac{\rho(1 - \sigma^2_{\mathbf{\widetilde{h}}})}{1 + \rho\sigma^2_{\mathbf{\widetilde{h}}}}. 
%\end{eqnarray} 

Channel capacity is defined as the maximal mutual information between known and observed signals ($x_t$, $y_t$, $y_d$) and the transmitted signal ($x_d$) over transmit signal distribution ($P_{x_d}$). \cite{shannon}

\begin{eqnarray}\label{cShannon}
C &:=& \sup_{P_{x_d}} \dfrac{1}{N} I(x_t, y_t, y_d ; x_d)\\
\end{eqnarray}

Applying chain rule for mutual information we have

\begin{eqnarray}\label{chain}
I(x_t, y_t, y_d ; x_d) &=& I(x_d; y_d | x_t, y_t) + I(x_d ; x_t, y_t)\\
&=& I(x_d; y_d | x_t, y_t)
\end{eqnarray} 
where $I(x_d ; x_t, y_t) = 0$, since data symbols ($x_d$) are independent of choosing training symbols and their corresponding output ($x_t$, $y_t$). Therefore the capacity is the maximal mutual information between transmitted and received data symbols ($x_d$, $y_d$) given transmitted training symbols and their corresponding received symbols ($x_t$, $y_t$).

\begin{eqnarray}\label{cChained}
C &:=& \sup_{P_{s_d}} \dfrac{1}{N} I(x_d; y_d | x_t, y_t)
\end{eqnarray}

The capacity depends upon the conditional distribution of given $y_t$ and $x_t$. For receiver structures that shape a particular estimation of the channel $\widehat{h}$, as long as information isn't "discarded" during the process, it is conceivable to achieve limit as given in (\ref{cChained}). However, a couple of data transmission systems that use training data do dispose of information in light of the fact that they frame an unequivocal estimation of the channel $\widehat{h}$ and use it as though it were right. The technique exhibited in this chapter to find a lower bound for capacity $C$ processes an unequivocal estimation of the channel $\widehat{h}$, reglates the estimation error to the added gaussian noise, and a short time later considers just the correlation between the transmitted signal and the subsequent noise. We, by then, get a lower bound by supplanting the subsequent noise by the most pessimistic situation with this same correlation. Under the assumption that $\widehat{h}$ is the conditional mean of $h$ (which is the estimate derived by MMSE estimator), given $y_t$ and $x_t$. We may now write (during the data transmission phase) that

\begin{eqnarray}\label{chModelTrPh}
y_d &=& \sqrt{\rho} x_d \widehat{h} + \sqrt{\rho}x_d \widetilde{h} + w_d
\end{eqnarray}

where $\widetilde{h} = h - \widehat{h}$ and we define the combined noise of AWGN and channel estimation error as $w_d = \sqrt{\rho}x_d \widetilde{h} + w$. Finding the capacity of the known-channel framework expects us to look at the most pessimistic impact that the added noise can have amid data transmission. We accordingly wish to find

\begin{eqnarray}
C \geq C_{worst} &=& \inf_{P_{w_d}} \sup_{P_{x_d}} I(y_d; x_d | \widehat{h}) \\
&=& \min_{\sigma^2_{w_d}} \max_{\sigma^2_{x_d}} \mathbb{E}\left[ \log (1 + \dfrac{\sigma^2_{x_d} |\widehat{h}|^2}{\sigma^2_{w_d}} ) \right] \\
&=& \min_{\sigma^2_{w_d}} \max_{\sigma^2_{x_d}} (1 - \alpha) \mathbb{E} \left[ \log(1 + \dfrac{\rho \sigma^2_{\widehat{h}}}{1 + \rho \sigma^2_{\widetilde{h}}}) |h|^2 \right]
\end{eqnarray}

The ratio
\begin{equation}
\rho_{\textit{eff}} =  \dfrac{\rho(1 - \sigma^2_{\mathbf{\widetilde{h}}})}{1 + \rho\sigma^2_{\mathbf{\widetilde{h}}}}. 
\end{equation}
can, therefore, be considered as an effective SNR. This bound does not require to be Gaussian.

%%%%%%%%%%%%%%%%%%%%%%  Hasibi COPY END  %%%%%%%%%%%%%%%%%%%%%%%

%%%%%%%%%%%%%%%% HASSINBI END %%%%%%%%%%%%%%%%%%%%%%%%%%%%

%%%%%%%%%%%%%%%% POLY BEGIN %%%%%%%%%%%%%%%%%%%%%%%%%%%%

\section{Short Packet Data Optimization Metric}

Consider a codingl model as described in the first section. At the receiver side, an error occurs if the decoder estimated message differs from the original message. It is desirable to find an encoder-decoder pair for some required error probability ($\epsilon$) with the minimum blocklength $n$ possible. Such goal can be represented as selecting $M$, $n-string$ messages, and their sets decoding pairs in the $\mathcal{A}^n$ space, such that after transmitting the original message, the corresponding estimated message after decoder can capture the original information with probability of at least $1-\epsilon$. Therefore, the ratio $\log M/n$ is known as \textit{rate} (with unit "bits/channel use"). 
\begin{equation}
R = \dfrac{\log M}{n}
\end{equation}

%\subsubsection{Channel Capacity}

\subsection{Traditional Optimization Metric: Ergodic Capacity}

\begin{figure}[H]
	\centering
	\includegraphics[width = 9cm]{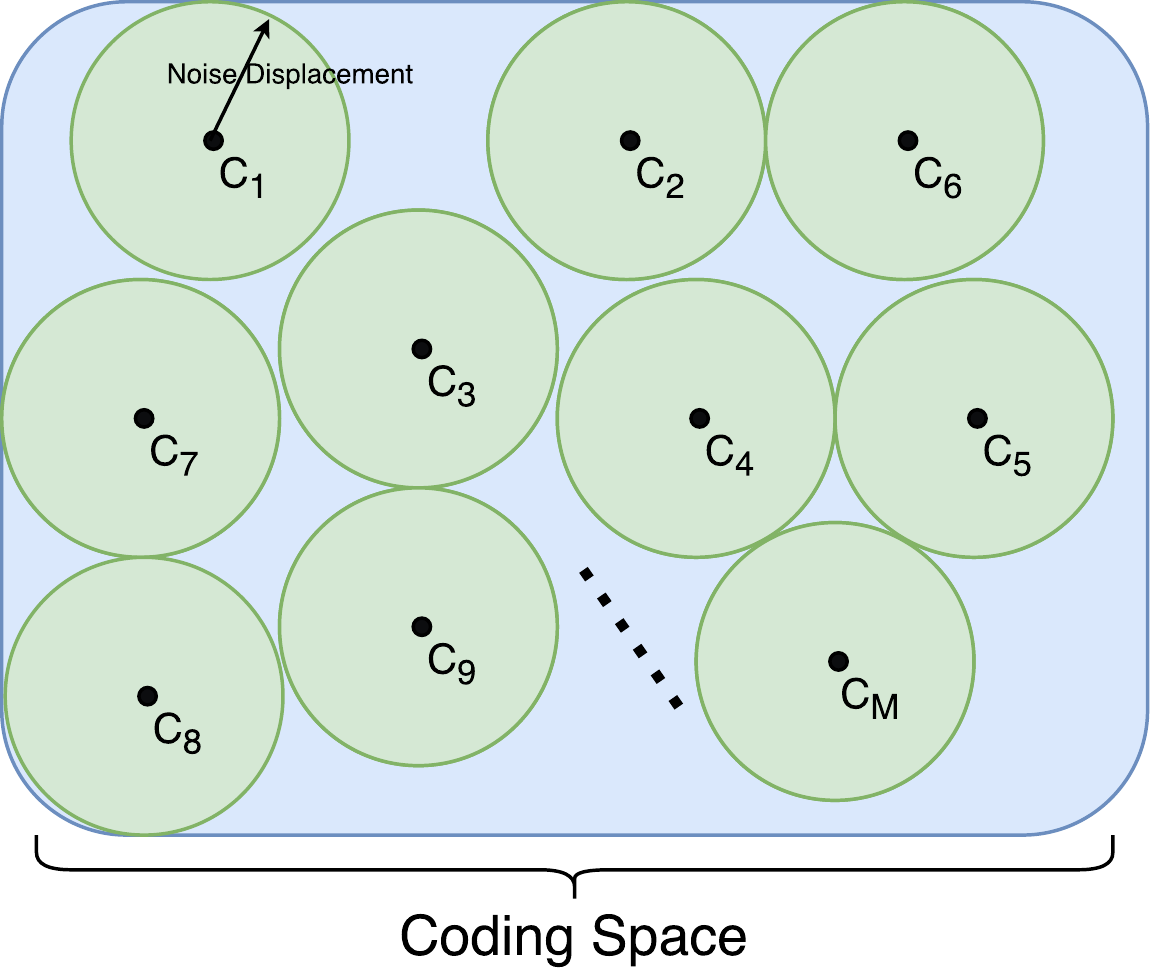}
	
	\caption{Packing messages (balls) in coding space.}
	\label{fig:balls}
\end{figure}

As mentioned before, the traditional optimization metric for training data size optimization is ergodic capacity. Shannon states in \cite{shannon} that there exist a ($n$, $M_n$, $\epsilon$) code with increasing blocklength which can achieve the communication rate of
\begin{equation}
R = \lim_{n\to\infty} \dfrac{1}{n} \log M_n > 0
\end{equation}
with vanishing probability of error
\begin{equation}
\lim_{n\to\infty}\epsilon_n \to 0
\end{equation}
Lets fix $n$ and $0 < \epsilon < 1$ and define the function
\begin{equation}
M^*(n, \epsilon) = \max\{M : \exists(n, M, \epsilon) codes\}
\end{equation}
which is the solution to the desired problem. $M^*(n, \epsilon)$ can be explained as maximum number of $n-string$ messages, "balls" that can be packed into the space $\mathcal{B}^n$; each of them should be able to capture the information with the probability $1-\epsilon$ when the center of the "ball" is being transmitted (as depicted in Fig. \ref{fig:balls}). Even for the simplest cases, finding the precise value of $M^*(n, \epsilon)$ is almost impossible. Shannon results states that independant of value of $\epsilon$
\begin{equation}
\label{C}
R = \limsup_{n\to\infty}\dfrac{1}{n} \log M^*(n, \epsilon) \leq C := \max_{P_X(x)} I(X;Y)
\end{equation}

where $P_X(x)$ is the probability distribution over original message set at the transmitter, %in tarif o check kon
$I(X, Y)$ is the mutual information between transmit random variable $X$ and receive random variable $Y$ defined as
\begin{equation}
I(X, Y) := \textbf{E}_{P_{XY}}\left[ \log(\dfrac{P_{XY}(x, y)}{P_X(x)P_Y(y)})\right]
\end{equation}
$C$ is called capacity of the channel and it is the maximum possible rate. Eq. (\ref{C}) can be interpreted as: reliably transmiting $nC$ data symbols with $n$ use of channel is possible. As an example, in an AWGN channel with signal to noise ratio ($\rho$), the capacity $C_{AWGN}(.)$ is given by
\begin{equation}
\label{CAWGN}
C_{AWGN}(\rho) = \log(1 + \rho)
\end{equation}
Shannon capacity perfectly described communication channels in a mathematical form, the only problem with the result (\ref{C}) is that it does not take $\epsilon$ into consideration i.e. it does not say that for a fixed $\epsilon$, what is the best value for blocklength $n$.

\subsection{Proposed Optimization Metric: Maximal Achievable Rate}

% in bakhsh o kollan tajdid nazar kon
Traditionally, the problem of finding the dependence between optimal blocklength and a fixed probability of error has been associated with finding the channel reliability function. However, the reliability function results is only available for block-fading channels and even for block-fading channels, this approach provides inferior results. For a variety of channels, \cite{polydmc} shows that the best approach to tackle this problem is through defining the channel dispersion. For a fixed probability of error ($\epsilon$) and a set of channels with capacity $C$, the channel dispersion is defined as \cite{polydmc}
\begin{equation}\label{Vdef}
V := \sup \dfrac{1}{n}\left( \dfrac{nC - \log M^*(n,\epsilon)}{Q^{-1}(\epsilon)} \right)^2
\end{equation}
The idea behind defining channel dispersion is to achieve tight asymptotic expression for channel rate. From the definition of channel dispersion (\ref{Vdef}), one can easily find asymptotic maximum achievable rate for a channel with known $C$ and $V$.
\begin{equation}\label{asymrate}
R^*(n, \epsilon) := \dfrac{1}{n}\log M^*(n, \epsilon) \simeq C - \sqrt{\dfrac{V}{n}} Q^{-1}(\epsilon)
\end{equation}
where $Q^{-1}(x)$ is the inverse Q-function. As it is shown in Eq. (\ref{asymrate}), channel dispersion ($V$) and channel capacity ($C$) are fundamental elements of finding such asymptotic expression for maximum achievable rate. Note that Eq. (\ref{asymrate}) implies that there is a penalty on the rate that is proportional to $1/\sqrt{n}$ to maintain the desired probability of error $\epsilon$ at a length $n$.

\subsubsection{Channel Dispersion for Memoryless Channels}

Finding the precise channel dispersion ($V$) for a channel is an impossible task since it requires finding the exact $M^*(.)$. Polyanskiy in \cite{polydmc} has studied different memoryless channels and tried to find tight approximations for channel dispersion. They used different lower and upper bounds to tightly approximate the channel dispersion. They show that for a variety of memoryless channels, the channel dispersion can be asymptotically computed as 
\begin{equation}\label{dispermemless}
V = \textbf{Var}\left[ i_{X;Y}(X,Y)|X \right]
\end{equation}
where the distribution of the $X$ is the capacity achieving distribution. and
\begin{equation}\label{infodest}
i_{X;Y}(x,y) := \log\dfrac{dP_{XY}}{d(P_XP_Y)}(x, y) = \log\dfrac{dP_{Y|X=x}}{dP_Y}(y)
\end{equation}
Note that taking expectation from Eq. (\ref{infodest}) will result in channel capacity, then again, if we take a variance of it, we will end up with the channel dispersion.

\subsubsection{Extension of Channel Dispersion for Fading Channels}

Fading channels can be modelled as a stationary process $H_i$. The channel is modeled as a cyclostationary process $H'_i$ with the cycle $T$ for simplification purposes. Each cycle $\{ H'_1, H'_2, \dots , H'_T \}$ is independent from the other cycles and they are distributed as $\{ H_1, H_2, \dots , H_T\}$. Then channel dispersion results of $H'_i$ are extend for the case $T \rightarrow \infty$, so that they can be extended to fading channel $H_i$. Such a channel ($H'_i$) can be represented as
\begin{equation}\label{cyclo}
\textbf{y}_k = \textbf{h}'_k \odot \textbf{x}_k + \textbf{w}_k \quad k = 1, \dots, n
\end{equation}
where $n$ is the coding blocklength, $\textbf{y}_k = (Y_1, Y_2, \dots, Y_T)_k$, $\textbf{x}_k = (X_1, X_2, \dots, X_T)_k$, $\textbf{h}'_k = (H'_1, H'_2, \dots, H'_T)_k$ and $\textbf{w}_k \sim \mathcal{N}(0, \rho\textbf{I}_T)$ where $\rho$ is the SNR. Since each cycle is independent of other cycles,  Eq. (\ref{cyclo}) represents a memoryless channel. Therefore the channel dispersion can be represented using Eq. (\ref{dispermemless}). According to \cite{polycite7}, the input capacity achieving distribution for (\ref{cyclo}) is when $\textbf{x} \sim \mathcal{N}(0, \textbf{I}_T)$. using this distribution and (\ref{cyclo}) in (\ref{infodest}), it can be shown that the information density can be expressed as
\begin{equation}
i_{XY}(x,y) = \dfrac{1}{2} \sum_{i = 1}^T \log (1 + \rho (H'_i)^2) + \dfrac{(H'_i)^2 X_i^2 + 2H'_iX_iW_i - \rho (H'_i)^2 W_i^2}{1 + \rho (H'_i)^2}\log e
\end{equation}
using (\ref{dispermemless}), the dispersion can be represented as
\begin{equation}
V_T = \textbf{Var}\left[0.5\sum_{i = 1}^T \log (1 + \rho (H'_i)^2)\right] + T\dfrac{\log^2 e}{2} \left(1 - \textbf{E}^2\left[\dfrac{1}{1 + \rho (H'_i)^2}\right]\right).
\end{equation}
Then extending to the case $T\rightarrow\infty$
\begin{equation}\label{finalV}
\mathtt{V}(\rho) = \lim_{T \rightarrow\infty} \dfrac{V_T}{T} = \textbf{Var}[C(\rho H^2)] + \dfrac{\log^2 e}{2} \left( 1 - \textbf{E}^2\left[\dfrac{1}{1 + \rho H^2}\right] \right)
\end{equation}
%%%%%%%%%%%%%%%% POLY END %%%%%%%%%%%%%%%%%%%%%%%%%%%%

%%%%%%%%%%%%%%%% ME BEGIN %%%%%%%%%%%%%%%%%%%%%%%%%%%%

\section{Optimizing Training Data Size for Short Packet Communication}
In this section, we provide an approximation of the achievable rate of a point-to-point communication ($R_{\textit{Tr}}^*(n, \epsilon, \rho)$) when training symbols and MMSE estimator are used to extract CSI at the receiver. Contrary to the assumption in previous section, here the channel is not known to the receiver but estimated using training symbols. %We divide pilot assisted scheme into two phases:\\
%\textit{1. Training phase: } Here we can describe our channel model as:
%\begin{equation}
%y^t(i) = \sqrt{\rho}x^t(i)h(i) + w^t(i),
%\end{equation}   
%where $x^t(i)$ are the pilot symbols and $y^t(i)$ are corresponding output symbols.\\
%\textit{2. Data transmission phase: } Here we can describe our channel model as:
%\begin{equation}
%y^d(i) = \sqrt{\rho}x^d(i)h(i) + w^d(i),
%\end{equation}   
 %where $x^d(i)$ are data symbols and $y^t(i)$ are corresponding output symbols. 

 \subsection{Using Maximal Acheivable Rate as Optimization Metric}

 During the data transmission phase, after MMSE estimation channel can be rewritten as:
\begin{equation}
\label{chMMSE}
y(i) = \sqrt{\rho}x(i)\widehat{h}(i) + \sqrt{\rho}x(i)\widetilde{h}(i) + w(i),
\end{equation}
 where  the CSI, $\widehat{h}(i)$, is perfectly known at receiver. The main problem here is the fact that the noise  $\sqrt{\rho}x(i)\widetilde{h}(i) + w(i)$  includes the channel estimation error. So the noise is not necessarily independent from the transmitted signal nor Gaussian.  First, we assume MMSE estimator, so that $\widetilde{h}(i)$ and $\widehat{h}(i)$ are orthogonal. Then we consider Gaussian noise, following the same approach used in ~\cite{mmse3}. Using those assumptions,  the channel defined in Eq. (\ref{chMMSE}) became similar to the channel introduced in previous section, with effective SNR equal to 
\begin{eqnarray}\label{roeffective}
\rho_{\textit{eff}} &=&  \dfrac{\rho(1 - \sigma^2_{\mathbf{\widetilde{h}}})}{1 + \rho\sigma^2_{\mathbf{\widetilde{h}}}}. 
\end{eqnarray} 

Finally, we took into account the number of symbols dedicated to the training phase ($n_t$ symbols don't carry data) to approximate  the achievable rate as: 
\begin{equation}
\label{32}
R_{\textit{Tr}}^*(n, \epsilon, \rho_{\textit{eff}}) \approx (1 - \alpha)\mathds{C}(\rho_{\textit{eff}}) - Q^{-1}(\epsilon)\sqrt{\dfrac{(1 - \alpha)\mathds{V}(\rho_{\textit{eff}})}{n}}.
\end{equation}
where efficient SNR can be derived using Eq. (\ref{roeffective})  and 
\begin{eqnarray}
\label{eq: 8}
\mathds{C}(\rho) = \log_2(e) e^{1/\rho} E_1(\dfrac{1}{\rho}), 
\end{eqnarray}
and $E_1(x) = \int_1^\infty t^{-1}e^{-xt} dt$ is the exponential integral. The channel dispersion $\mathds{V}(\rho)$ can be derived as shown in Eq. (\ref{finalV}):

\subsection{Numerical Results}
In this section, we numerically evaluate the optimal pilot overhead for ultra-reliable short-packet transmission. These numerical results are derived by solving the derivative of Eq. (\ref{32}) w.r.t $\alpha$ equal to zero.  For comparison purpose we also optimize the training data using the ergodic capacity  \cite{lozano2008}.   
These comparisons are shown in Fig. \ref{both_eps}-\ref{Ropt_block}.  Our simulation results prove that our optimization approach will result in increase of around 10\% in optimal rate for short packet communication.  We performed simulations for a broad range of variables such as:

\subsubsection{Optimal Training Data Size vs Probability of Error} 
\begin{figure}[H]
	\centering
	\includegraphics[width=9.5cm]{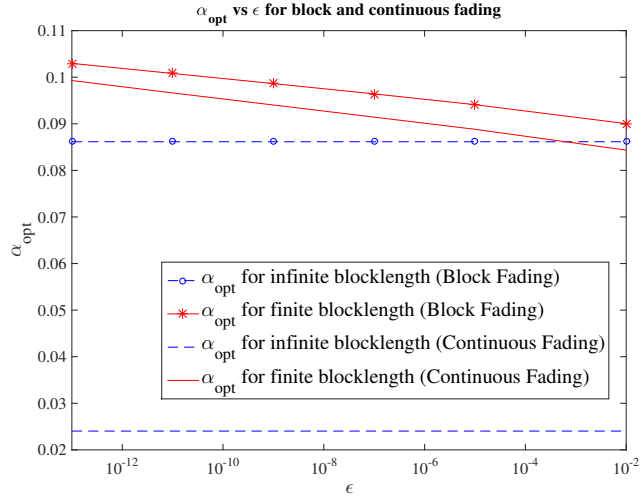}
	\vspace{-0.4cm} \caption{Optimal pilot overhead for infinite and finite blocklength in block and continuous fading model vs. $\epsilon$ with $n = 30$ and SNR = 15dB.}	
	\label{both_eps}
\end{figure}

We first compare the optimal pilot overhead for different error probabilities. The difference between our optimal pilot overhead values and those derived using ergodic capacity increases with decreasing probability of error. Thus, it is  very important to use the new formulation for ultra-reliable communication systems. We have similar results for both block and continuous fading (Fig. \ref{both_eps}).

\subsubsection{Optimal Training Data Size vs Blocklength}  
\begin{figure}[H]
	\centering
	\includegraphics[width=8.5cm]{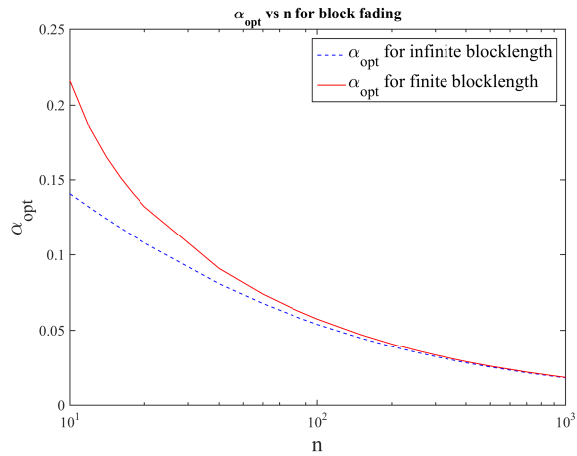}
	\vspace{-0.4cm}\caption{Optimal pilot overhead for infinite and finite blocklength in block fading model vs. Blocklength with SNR = 8dB and $\epsilon$ = 1e-9.}
	\label{block_n}
\end{figure}

\begin{figure}[H]
	\centering
	\includegraphics[width=8.5cm]{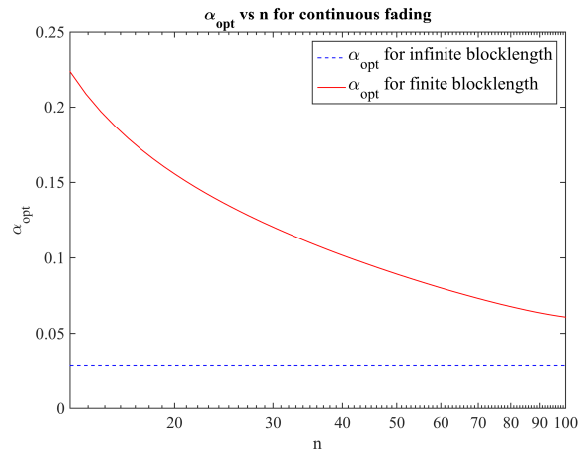}
	\vspace{-0.4cm}\caption{Optimal pilot overhead for infinite and finite blocklength in continuous fading model vs. Blocklength with SNR = 23dB, $\epsilon$ = 1e-9 and $f_D$ = 0.02.}	
	\label{cont_n}
\end{figure}

We numerically evaluated the optimal pilot overhead for different blocklength. We considered both block and continuous fading. Note that when we use ergodic capacity optimization, the blocklength is assumed infinite but the variance of the channel estimation error varies with $n$ for block fading.   
As shown in Fig. \ref{block_n}, the difference is higher in small blocklength. This suggests that our approach is more adequate to short packet transmission. Moreover, we can see in Fig. \ref{cont_n} that the pilot overhead increases with small blocklength when we need more and more pilot symbols to compensate the channel estimation mismatch.

\subsubsection{Optimal Training Data Size vs SNR and  Normalized Doppler Frequency}  
\begin{figure}[H]
	\centering
	\includegraphics[width=9.5cm]{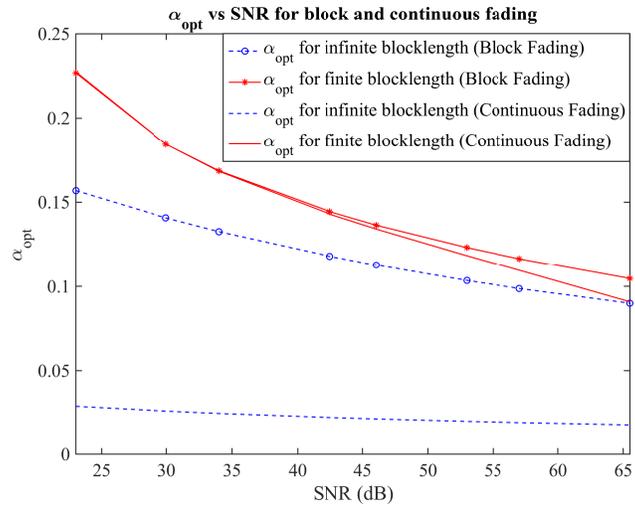}
	\vspace{-0.4cm}\caption{Optimal pilot overhead for infinite and finite blocklength in block and continuous fading model vs. SNR with $n = 40$ and $\epsilon$ = 1e-9.}	
	\label{both_SNR}
\end{figure}

\begin{figure}[H]
	\centering
	\includegraphics[width=9.5cm]{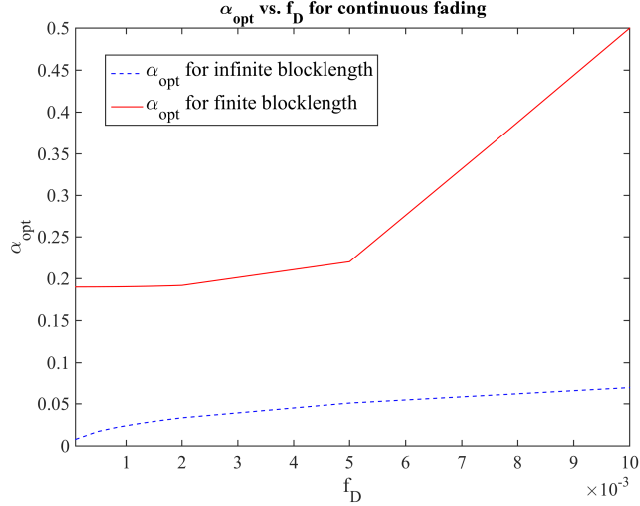}
	\vspace{-0.4cm}\caption{Optimal pilot overhead for infinite and finite blocklength in continuous fading model vs. $f_D$ with $n = 10$, SNR = 16dB and $\epsilon$ = 1e-9.}	
	\label{cont_fd}
\end{figure}

The simulations, illustrated in Fig. \ref{both_SNR} and Fig. \ref{cont_fd},  show that the optimal pilot overhead decreases with SNR and increases with $f_D$, as expected. In addition,  the difference between our optimal pilot overhead values and those derived using ergodic capacity  is greater at low SNR and high normalized Doppler frequency. 


\subsubsection{Effect of Optimal Training Data Size on Optimal Rate} 
\begin{figure}[H]
	\centering
	\includegraphics[width=7.7cm]{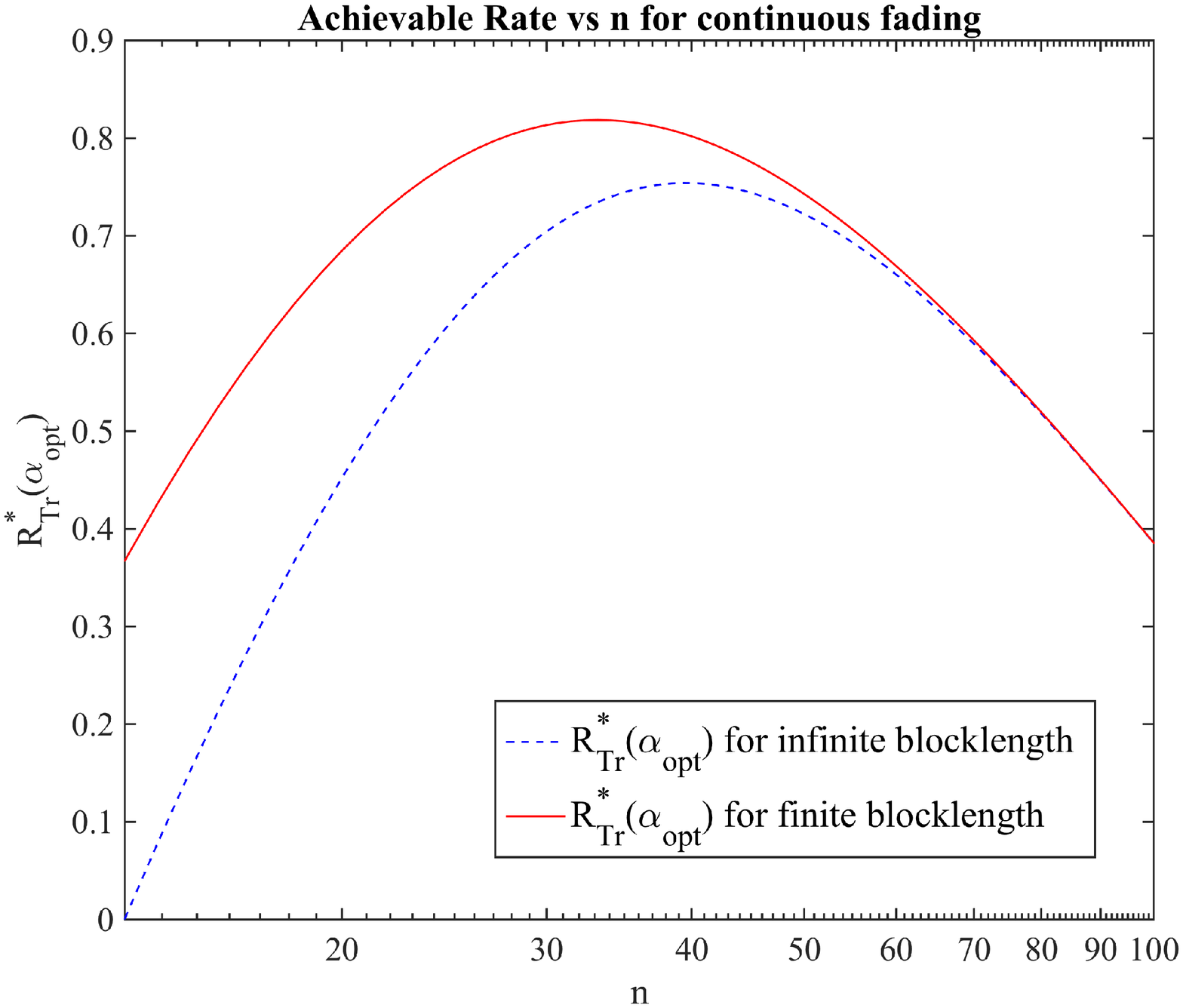}
	\vspace{-0.4cm}\caption{Acheivable rate using infinite and finite blocklength $\alpha_{opt}$ in continuous fading model vs. $n$ with SNR = 20dB, $\epsilon$ = 1e-12 and $f_D$ = 0.02.}	
	\label{Ropt_cont}
\end{figure}

\begin{figure}[H]
	\centering
	\includegraphics[width=7.7cm]{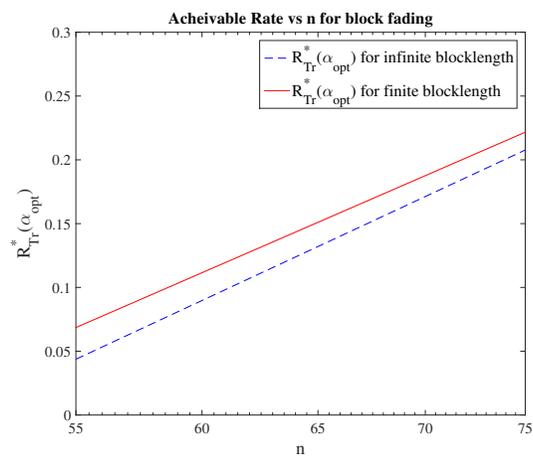}
	\vspace{-0.4cm}\caption{Achievable rate using infinite and finite blocklength $\alpha_{opt}$ in block fading model vs. $n$ with SNR = 7dB, $\epsilon$ = 1e-9 and $f_D$ = 0.02.}	
	\label{Ropt_block}
\end{figure}

We evaluated the rate at the optimum values of $\alpha$ evaluated in this paper and optimum alpha for infinite blocklength. As illustrated in Fig. \ref{Ropt_block} our optimization approach will result in increase of roughly 10\% in rate with block fading. Fig. \ref{Ropt_cont} shows even more significant increase in rate with continuous fading. Simulations show that our approach always results in a higher rate. Fig. \ref{Ropt_cont} also shows that due to the channel estimation mismatch in continuous fading, increasing blocklength after a certain blocklength ($n = 29$ in this case) results in decreasing rate.

\section{Conclusion}

The objective of this chapter is to enhance the packet efficiency by optimizing the training data size for ultra-reliable short packetl transmission. We considered a point-to-point communication model in which one node sends messages to a central node, or base-station, over AWGN channel with Rayleigh fading channel. We formalized the optimization problem as far as surmised achievable rates as capacity of square length, pilot length, and mistake likelihood. Simulation resluts demonstrated that it is critical to consider the training data size and the error probability while optimizing the training data size.

%%%%%%%%%%%%%%% MOVE THIS %%%%%%%%%%%%%%%%%%%%

%{\color{red} MOVE ALL AFTER HERE}
%
%
%
%
%\section*{Acknowledgment}
%Funding for this research was partially supported by NSF under Award Number 1620794.

%\myPart{II}{Models}{4cm}

% Chapter 4

\chapter{Training Data Design for Integrated Communication and Sensing} % Main chapter title

\label{Chapter4} % For referencing the chapter elsewhere, use \ref{Chapter4} 

\begin{figure*}
	\centering
	%\subfigure[Transmit]{	\includegraphics[scale=0.35]{1.pdf}
	%}
	%\subfigure[Receive]{	\includegraphics[scale=0.35]{2.pdf}
	%}
	\includegraphics[width=16cm]{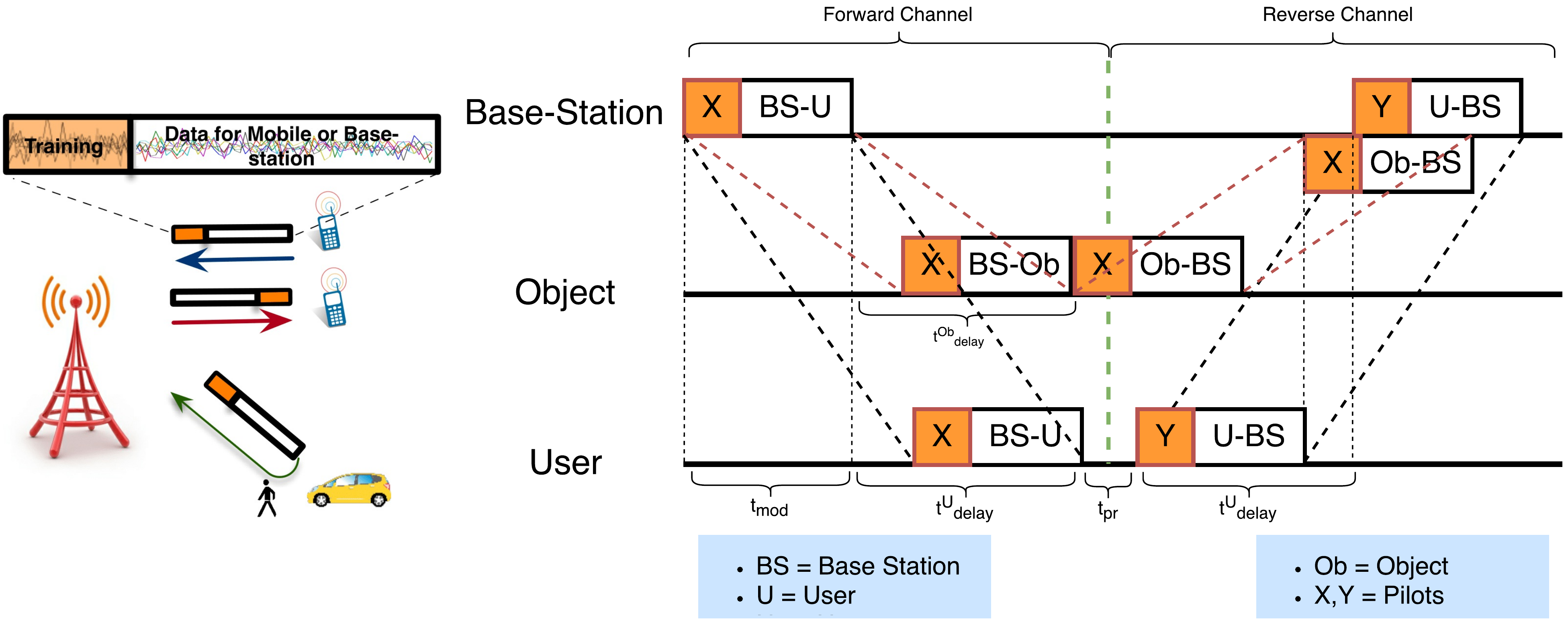}
	
	\caption{The \emph{ComSens} communication and sensing integration framework.}	
	\label{comb}
\end{figure*}

Our second proposed solution to increase spectral efficiency has been briefly introduced in Chapters \ref{Chapter1} and \ref{Chapter2}. As mentioned before, the idea is to design a multifunction radar and communication system which integrates sensing abilities into an already existing communication system. In this chapter, this idea will be explained in detail, the proposed scheme will be explained and analyzed, the design algorithm will be discussed, and the performance of the designed system will be evaluated through simulations.

\section{Fusion of Communication and Radar Operations}

The problem settings and the proposed scheme details are described in this section.

\subsection{The Proposed Integration Scheme}
The ComSens framework operates by exploiting the two-way communication between the end-users and base-station. Before discussing the issue of designing the training data sequences, we will initially address how the base-station and end-client trade messages and sense the environment simultaneously and over a similar spectrum.  We consider a model with $M$-user multiple-access-broadcast channel (MABC) as depicted in Fig. \ref{comb}. Note that such a system model, in which several end-users wish to trade messages with a base-station, is a model that catches the behavior of present and future cell systems. We assume  half-duplex end-user nodes that may transmit or receive at a given time, on a given frequency, but not both, leading to the need to describe protocols, or which nodes transmit when.  We consider time division duplex (TDD) two-way system as duplex scheme and for multiple access both FDMA and TDMA can be used.  %The transmission time is divided into $M$ time periods, each of which consist of two phases. During the first period, the terminal nodes transmit to the base-station. During the second period the base-station transmits  to the terminal nodes; see Fig. \ref{comb}.
For each user, time is devided into \textit{forward channel} and \textit{reverse channel} (as in TDD scheme). During the former time, Base-station transmits the packet and during the latter user transmits the packet. 
The base-station (BS) sends a packet $s$ to the end-user U.  The end-user can extract its own message after channel estimation using the downlink training data -- labeled $\boldsymbol{X}$. Contrary to most of the current works on integrated radar and communication systems, the data transmission proposed here is similar to the conventional half-duplex transmission.  This guarantees a high-data rate to efficiently accommodate downlink  traffic. In the meantime, the packet $s$ is reflected from objects in the area. The base-station watches the echo of its own transmit sequence, and distinguishes the objects and their distance and relative speed. With ComSens, the base-station jointly estimates the radar return and extract the uplink message from end-user U after channel estimation using uplink training data -- labeled $\boldsymbol{Y}$.  The principal constraint in the performance of radar sensing is the simultaneous reception of the radar echo and uplink packet. Therefore the main goal of this work is to design the uplink and downlink training data sequences.  We design the two training data sequences  to be uncorrelated to each other so that they can be distinguished from each other at the base-station. After separation of two training data sequences, the base-station uses the packet with uplink training data for communication purposes and the reflected downlink packet for sensing.

%{\color{red}The text we had before (to use in modifying the text below): We partition the training data signals into two parts; each part to be transmitted by a communication agent. Therefore, after an initial training data signal transmission, the receiver communicates a different signal coming back to the original device of the base station, while a passive object will back-scatter the same signal--- thus paving the way for the system to distinguish various signals, and consequently, being able to perform as a radar system at the same.}

 %As it is shown in Fig.~\ref{protocol}, ComSens consists of a base station, a communication user and an object. First, base station starts by transmitting packets with training data 1. Mobile user receives the packet with training data 1, and transmits another packet with training data 2. The object will back-scatter the same packet with training data 1. Base station receives a packet with training data 2 from the user and  a packet with training data 1 from the object. 

\emph{Remark:} From the above discussion it must be clear that ComSens uplink communications may be subject to interference from the radar echo.  Note that the echo signal is received at the base-station with high attenuation due to the two-way
link  (from the base-station to the object and from the object to the base-station) and the absorption at the object so its impact on the uplink communication is negligible.

\subsection{Time and Range Analysis}

Here we analyze the timing for the proposed protocol and discuss the limitations on the range of the objects that the sensing system is able to detect. Then we evaluate the limitations in a practical scenario to get a better sense of them. 

%\begin{figure}
%\centering
%	\includegraphics[width=12cm]{Figures/ch5/Diagram2.pdf}
%
%	\caption{Time Diagram.}	
%	\label{diagram}
%\end{figure}

\subsubsection{Limitations}

Consider one TDD frame for an end-user U (as it is shown in Fig. \ref{comb}). At the forward channel time, base-station transmits the packet. End-user receives the packet at $t_{mod}+t^U_{delay}$ where $t_{mod}$ is the modulation and transmission time and $t^U{delay}$ is the propagation time between end-user and the base-station. Packet is processed at the end-user in $t_{pr}$ time. Then, end-user transmits the packet in the reverse channel time and base-station receives it at the time $t_1 = t_{mod}+2t^U_{delay}+t_{pr}$. On the other hand, transmitted packet from the base-station is also received at the object at the time $t_{mod} + t^{Ob}_{delay}$ where $t^{Ob}_{delay}$ is the propagation time between object and the base-station. The packet is then back-scattered from the object and received at the base-station at the time $t_2 = t_{mod} + 2t^{Ob}_{delay}$.
We design downlink training sequence  and uplink training sequence to be uncorrelated to each other for $k$ time lags. Therefore, if two received signals (from user and object) have arrival time difference ($t_2-t_1$) of at most $k$, they are distinguishable from each other. On the other hand, if $t_2-t_1 > k$, the radar signal cannot be recognized and it will be considered as weak interference for communication system. Consequently, our proposed integrated radar system will perform when $t_2-t_1 \leq k$. Substituting $t_1$ and $t_2$ we have:

\begin{eqnarray}
\label{RadCond}
t^{U}_{delay} - t^{Ob}_{delay} \leq \dfrac{t_{pr}+k}{2}
\end{eqnarray}
where
\begin{equation}
\label{del1}
t^{Ob,U}_{delay} = \dfrac{d^{Ob,U}}{\nu T_s}
\end{equation}
%\begin{equation}
%\label{del2}
%t_{delay2} = \dfrac{d_{2}}{\nu T}
%\end{equation}
and $d^{U}$ and $d^{Ob}$ are respectively the distance of user and the object from the base-station, $T_s$ is symbol time in our system and $\nu$ is the speed of electromagnetic wave in the space. Using Eq. (\ref{RadCond}, \ref{del1}) we have

\begin{equation}
\label{notrange}
 d^{Ob} \leq  d^{U} + \dfrac{\nu T (t_{pr} + k)}{2} 
\end{equation}
%if $d_2$ is less than the lower bound 

%Assume that we have $N$ symbols to transmit (since we have a MIMO system, here $N = N_s/n_t$ where $N_s$ is number of symbols). Transmission schedule for our system is shown in Fig. \ref{comb}. Packet sent from base-station to end-user 1 is received at the time $N + t_{delay1}$ (the process of transmission takes $N$ blocks of time and we define $t_{delay1}$ as the time distance between base-station and end-user 1). This transmitted packet is also received at the object at the time $N + t_{delay2}$ ($t_{delay2}$ is defined as the time distance between base-station and object). Now packet received at end-user 1 is processed (we assume a processing time of $t_p$) and then a packet is transmitted back to base-station. This packet is received at base-station at the time $N + 2t_{delay1} + t_p$, while the back-scattered packet (from the object) is received at the base-station at the time $N + 2t_{delay2}$. Then, the same scenario happens to end-user 2.  

\subsubsection{Practical Scenario}

Communication cell towers have a range between $35km$ to $72km$. We consider our user to be (as a medium distance) at the distance $d^{U} = 25km$ of the base-station. %As it is shown in Fig. \ref{range}, for a fixed user, we have a blind side for radar (white region) and a working radar region (green region).
 Assume that the symbol time $T_s = 25 \mu s$  and processing time $t_{pr} = T_s$ where speed of electromagnetic wave is $\nu = 3\times10^{8}$, assuming we design our training sequences to be uncorrelated for $k = 4$. Such a system would have a radar range of 43.75km ($d^{Ob} \leq 43.75km$).

\subsection{Channel Model}
We consider the same settings as in \cite{soltan}. More precisely, we consider a point-to-point block fading narrowband MIMO channel where transmit and receive antennas are represented with $n_T$ and $n_R$ respectively. Deffining $\boldsymbol{P} \in \mathcal{C}^{B\times n_T}$ as a matrix where the rows are the training sequence at every transmitter antenna. At the training stage, channel can be represented as

\begin{equation}
\boldsymbol{Y} = \boldsymbol{H}\boldsymbol{P}^T + \boldsymbol{N}
\end{equation}  
where $\boldsymbol{Y} \in \mathcal{C}^{n_R\times B}$ is the received sequence, $\boldsymbol{H} \in \mathcal{C}^{n_R\times n_T}$ is the MIMO channel when $\boldsymbol{H}(i,j)$ denotes the MIMO channel gain between $i^{th}$ transmitter and $j^{th}$ receiver and $\boldsymbol{N} \in \mathcal{C}^{n_R\times B}$ is the noise matrix. We assume Gaussian noise i.e. $vec(\boldsymbol{N}) \sim \mathcal{CN}(\textbf{0}, \boldsymbol{M}) $ where $\boldsymbol{M} \in \mathcal{C}^{Bn_R\times Bn_R}$ denotes noise covariance matrix. We also assume $vec(\boldsymbol{H}) \sim \mathcal{CN}(\textbf{0}, \boldsymbol{R})$ where $\boldsymbol{R} \in \mathcal{C}^{n_Tn_R\times n_Tn_R}$ denotes channel covariance matrix.

\begin{figure}
\centering
	\includegraphics[width=5cm]{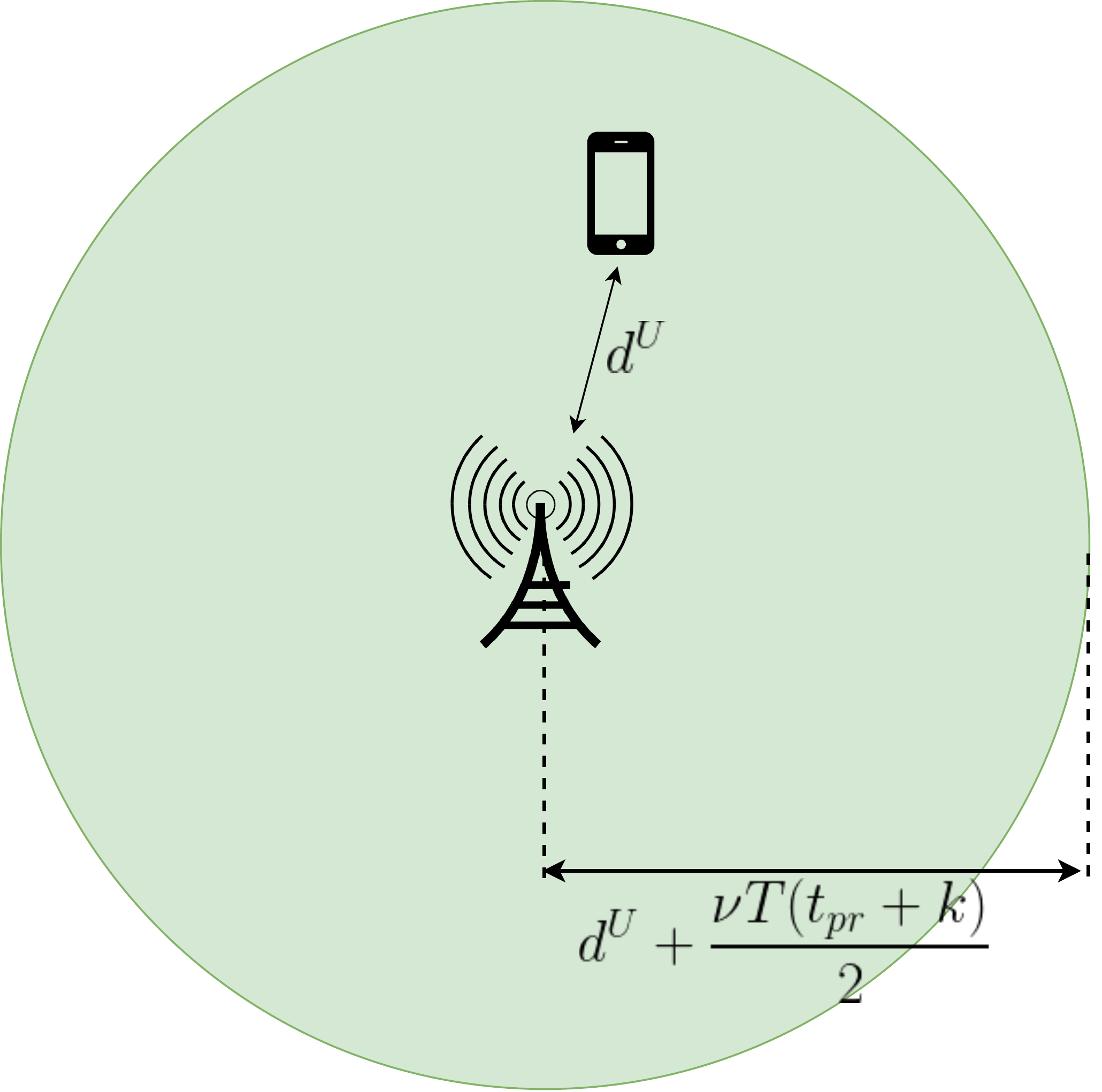}

	\caption{The radar operational range of ComSens.}	
	\label{range}
\end{figure}

\section{Training Sequence Design}

In this section, we design the training coefficients gathered in the matrix $\boldsymbol{P}$--to acquire an accurate channe estimate l $\boldsymbol{H}$--while simultaneously satisfying a set of radar performance criteria. For an accurate channel estimation, one may resort to a  minimization of the channel mean-squared error (MSE), expressed as \cite{soltan}
\begin{eqnarray} \label{eq:MSE}
MSE=%tr\left[\theta\right] \nonumber\\ 
tr\left[ \left(\boldsymbol{R}^{-1} + (\boldsymbol{P}\otimes \boldsymbol{I}_{n_R})^H \boldsymbol{M}^{-1}(\boldsymbol{P}\otimes \boldsymbol{I}_{n_R}) \right)^{-1} \right].
\end{eqnarray}
Let $\widetilde{\boldsymbol{P}}  \triangleq {\boldsymbol{f} P} \otimes {\boldsymbol{f} I}_{n_{R}} \in \mathbb{C}^{B n_R  \times n_T n_R}$, and by utilizing matrix inversion lemma we have
\begin{eqnarray}
\boldsymbol{\theta} &\triangleq& \left( \boldsymbol{R}^{-1} + \widetilde{\boldsymbol{P}}^H \boldsymbol{M}^{-1} \widetilde{\boldsymbol{P}} \right)^{-1} \\  &=& \boldsymbol{R} - \boldsymbol{R} \widetilde{\boldsymbol{P}}^H \left( \boldsymbol{M} + \widetilde{\boldsymbol{P}} \boldsymbol{R} \widetilde{\boldsymbol{P}}^H \right)^{-1} \widetilde{\boldsymbol{P}} \boldsymbol{R} ,  % \nonumber
\end{eqnarray}
where $MSE = tr[\theta]$. Now let
\begin{equation}\label{eq:def1}
\boldsymbol{Q} \triangleq \left(
\begin{array}{cc}
\boldsymbol{R} & \boldsymbol{R}  \widetilde{\boldsymbol{P}}^H \\%\nonumber
\widetilde{\boldsymbol{P}} \boldsymbol{R} & \boldsymbol{M} + \widetilde{\boldsymbol{P}} \boldsymbol{R} \widetilde{\boldsymbol{P}}^H
\end{array}\right) \in \mathbb{C}^{(B+n_T)n_R  \times (B+n_T)n_R},
\end{equation}
\begin{eqnarray}
\boldsymbol{U} \triangleq (\boldsymbol{I}_{n_T n_R}\;\; \boldsymbol{0}_{n_T n_R \times B n_R})^T \in \mathbb{C}^{(B+n_T)n_R  \times n_T n_R},
\end{eqnarray}
and observe that \cite{petersen2008matrix},
\begin{equation}\label{eq:def2}
\boldsymbol{U}^H \boldsymbol{Q}^{-1} \boldsymbol{U}=\boldsymbol{\theta}^{-1}.
\end{equation}
In light of the above, the authors in \cite{soltan} propose a cyclic optimization approach to minimizing the MSE in \eqref{eq:MSE}: Consider an auxiliary variable $\boldsymbol{V} \in \mathcal{C}^{n_Tn_R \times Bn_R}$ such that
\begin{equation}
\label{fvx}
F(\boldsymbol{V}, \boldsymbol{P}) := tr\left[\boldsymbol{V}^H\boldsymbol{Q}\boldsymbol{V}\right].
\end{equation}
 The minimizer $\boldsymbol{V}$ of (\ref{fvx}) can be obtained as \cite[p. 354]{peterbook-spectral}

\begin{eqnarray}
\label{minimizer}
\boldsymbol{V}_* =
\begin{pmatrix}
\boldsymbol{I}_{n_Tn_R} \\
-\left(\boldsymbol{M} + \widetilde{\boldsymbol{P}}\boldsymbol{R}\widetilde{\boldsymbol{P}}^H \right)^{-1}\widetilde{\boldsymbol{P}}\boldsymbol{R}
\end{pmatrix}
\end{eqnarray}
By substituting (\ref{minimizer}) in (\ref{fvx}), one can verify that
\begin{equation}
F(\boldsymbol{V}_*, \boldsymbol{P}) = tr\left[\boldsymbol{\theta}\right] = MSE.
\end{equation}
Therefore, in order to optimize the MSE we can use a cyclic optimization of (\ref{fvx}) with respect to $\boldsymbol{V}$ and $\boldsymbol{P}$. In particular, it was shown in \cite{soltan} that the optimization of  (\ref{fvx}) with respect to  $\boldsymbol{P}$ can be cast at each (cyclic) iteration as:

\vspace{0.03cm}
\begin{equation}
\label{Piteration}
\min_{\boldsymbol{P}^{h+1} \in \Omega} \left|\left| \boldsymbol{P}^{(h+1)} - \boldsymbol{P}_{\Sigma}^{(h)} \right|\right|^2_{2},
\end{equation} 
\vspace{0.03cm}

\noindent where $\boldsymbol{P}_{\Sigma}^{(h)}$  is constructed from $\boldsymbol{P}^{(h)}$ at each iteration (see \cite{soltan} for details). For the two-part training sequence employed in ComSens,  define:
\begin{eqnarray}
\boldsymbol{P}_{DL} &:=& \boldsymbol{X} \\
\boldsymbol{P}_{UL} &:=& \boldsymbol{Y}
\end{eqnarray}
\vspace{0.03cm}

\noindent where $\boldsymbol{X} \in \mathcal{C}^{B\times n_T}$ is the downlink training sequence  contributing at both radar and communication modes and $\boldsymbol{Y} \in \mathcal{C}^{B\times n_R}$ is the uplink training sequence which contributes only in communication mode. Thus, (\ref{Piteration}) becomes

\vspace{0.03cm}
\begin{eqnarray}
\label{optimiz}
\min_{\boldsymbol{x}, \boldsymbol{Y} \in \Omega} \left|\left| \boldsymbol{X} - \boldsymbol{X}_{\Sigma}\right|\right|_2 ^2+ \left|\left| \boldsymbol{Y} - \boldsymbol{Y}_{\Sigma}\right|\right|_2^2,
\end{eqnarray}
\vspace{0.03cm}

\noindent where the constraint set $\Omega$ is yet to be defined. As indicated earlier, $\boldsymbol{X}$ and $\boldsymbol{Y}$ should have \emph{low correlation} with each other and $\boldsymbol{X}$ should have an \textit{impulse-like} autocorrelation. We describe the training constraints in three categories:
%\vspace{-1cm} 
\begin{enumerate}
\item Both training sequences should have fixed transmit powers given by
\begin{eqnarray}
 || \boldsymbol{x}_q ||_2^2 \leq p, \quad 1 \leq q \leq n_T\\
 || \boldsymbol{y}_l ||_2^2 \leq p, \quad 1 \leq l \leq n_R
\end{eqnarray}
where $\boldsymbol{x}_q$ and $\boldsymbol{y}_l$ are column vectors of $\boldsymbol{X}$ and $\boldsymbol{Y}$ and $p$ is the power upper-bound.
\item To resolve ambiguity between radar reflections and communication signals, training sequences (and their time lags up to $k$ lags) should be uncorrelated to each other; i.e their cross correlation must be zero or very small at least for a number of time lags (forming a zero correlation zone \cite{fan1999class}): \\
\begin{equation}
 \boldsymbol{X}^T \boldsymbol{J}_{i} \boldsymbol{Y} \simeq \boldsymbol{0}^{n_T\times n_R}, \quad 0 \leq i \leq k,
\end{equation}
where $\boldsymbol{J}_{k} \in C^{B \times B}$ is a shift matrix that shifts a matrix by $k$ time lags. Clearly $\boldsymbol{J}_0$ is identity matrix.
\item Radar training sequence should be impulse-like; i.e. its auto-correlation must be zero or very small at least for a number of time lags:\\
\begin{equation}
 \boldsymbol{X}^T \boldsymbol{J}_{i}\boldsymbol{X} \simeq \boldsymbol{0}^{n_T\times n_T},  \quad 0 \leq i \leq k.
\end{equation}
\end{enumerate}  

Consequently, one can solve the following optimization problem to design our training sequences:

\begin{eqnarray}
\label{OP}
\min_{\boldsymbol{X}, \boldsymbol{Y} \in \Omega}&& \left|\left| \boldsymbol{X} - \boldsymbol{X}_{\Sigma}\right|\right|_2 ^2+ \left|\left| \boldsymbol{Y} - \boldsymbol{Y}_{\Sigma}\right|\right|_2^2 \\
s.t. \quad  &&|| \boldsymbol{x}_q ||_2^2  \leq p, \quad 1 \leq q \leq n_T; \nonumber\\
 &&|| \boldsymbol{y}_l ||_2^2 \leq p,\quad 1 \leq l \leq n_R; \nonumber\\
 && \boldsymbol{x}_q^T \boldsymbol{J}_{i}\boldsymbol{y}_l\leq \epsilon,  \quad 1 \leq i \leq k; \nonumber\\
  &&\boldsymbol{x}_q^T \boldsymbol{J}_{i}\boldsymbol{x}_q\leq \epsilon,  \quad 1 \leq i \leq k; \nonumber
\end{eqnarray}
where $\epsilon$ is a very small number (in this paper we use $10^{-5}$) to achieve equality constraints. In order to tackle (\ref{OP}) we can use cyclic optimization\cite{stocia2004cyclic}. We define:

%In order to solve optimization problem Eq. (\ref{Convex}), we first make it an optimization problem with vector constraints instead of matrix constraints.
\begin{equation}
G(\boldsymbol{X},\boldsymbol{Y}) := \left|\left| \boldsymbol{X} - \boldsymbol{X}_{\Sigma}\right|\right|_2 ^2+ \left|\left| \boldsymbol{Y} - \boldsymbol{Y}_{\Sigma}\right|\right|_2^2
\end{equation} 
Then one can perform a cyclic procedure to minimize $G(\boldsymbol{X},\boldsymbol{Y})$ as follows: We start with an initial value $\boldsymbol{Y} = \boldsymbol{Y}^0$. Then we comupte $\boldsymbol{X}^i$ by tackling minimization problem in Eq. (\ref{XConv}) and $\boldsymbol{Y}^i$ by tackling minimization problem in Eq. (\ref{YConv}). %until a stop criterion ($||\boldsymbol{X}^{i}-\boldsymbol{X}^{i-1}||_2+||\boldsymbol{Y}^{i}-\boldsymbol{Y}^{i-1}||_2 < \zeta$ for some $\zeta > 0$) is satisfied. 
More precisely:

%sdadas sdasd asda sdadas sdasd asda sdadas sdasd asda sdadas sdasd asda sdadas sdasd asda sdadas sdasd asda sdadas sdasd asda sdadas sdasd asda
\begin{eqnarray}
\label{XConv}
\boldsymbol{X}^i &=& arg \min_{\boldsymbol{X}} G(\boldsymbol{X},\boldsymbol{Y}^{i-1}) \\
s.t. &&  || \boldsymbol{x}_q ||_2^2  \leq  p, \quad\quad\quad 1 \leq q \leq n_T; \nonumber\\
 && \boldsymbol{x}_q^T \boldsymbol{J}_{m}\boldsymbol{y}^{i-1}_l =  0,  \quad 1 \leq m \leq k; \nonumber\\
  &&\boldsymbol{x}_q^T \boldsymbol{J}_{m}\boldsymbol{x}_q\leq \epsilon,  \quad\quad 1 \leq m \leq k; \nonumber
\end{eqnarray}

\begin{eqnarray}
\label{YConv}
\boldsymbol{Y}^i &=& arg \min_{\boldsymbol{Y}} G(\boldsymbol{X}^i,\boldsymbol{Y}) \\
s.t. &&|| \boldsymbol{y}_l ||_2^2 \leq p, \quad\quad\quad  1 \leq l \leq n_T; \nonumber\\
&& (\boldsymbol{x}^i_q)^T \boldsymbol{J}_{m}\boldsymbol{y}_l\leq 0,  \quad 1 \leq m \leq k; \nonumber
\end{eqnarray}
%Moreover, constraint 3 is changed to make a convex constraint. Eq. (\ref{OP}) is now a convex optimization problem with convex constraints which can be solved using convex optimization techniques.
where $1 \leq q \leq n_T$ and $1 \leq l \leq n_R$. Note that since now the second constraint in both (\ref{XConv}) and (\ref{YConv}) are affine constraints, we replaced them with equality. Eq. (\ref{YConv}) is now a convex optimization problem and solvable using convex optimization. However, the third constraint in (\ref{XConv}) is not convex. We can rewrite Eq. (\ref{XConv}) in form:

\begin{eqnarray}
\label{XConv2}
\boldsymbol{X}^i &=& arg \min_{\boldsymbol{X}} G(\boldsymbol{X},\boldsymbol{Y}^{i-1}) \\
s.t. &&  || \boldsymbol{x}_q ||_2^2  \leq  p, \quad\quad\quad\quad\quad\quad\quad\quad\quad\quad  1 \leq q \leq n_T; \nonumber\\
 && \boldsymbol{x}_q^T \boldsymbol{J}_{m}\boldsymbol{y}^{i-1}_l =  0,  \quad\quad\quad\quad\quad\quad\quad\quad 1 \leq m \leq k; \nonumber\\
  &&\boldsymbol{x}_q^T (\boldsymbol{J}^T_{m}+\boldsymbol{J}_{m}+2\boldsymbol{I}_{m})\boldsymbol{x}_q\leq 2p,  \quad\quad 1 \leq m \leq k; \nonumber
\end{eqnarray}
the third constraint in (\ref{XConv2}) is now in quadratic convex form since $(\boldsymbol{J}^T_{m}+\boldsymbol{J}_{m}+2\boldsymbol{I}_{m})$ is a symmetric positive semi-definite matrix. Note that the optimization problem is still the same (since $\boldsymbol{x}_q^T \boldsymbol{J}_{m}\boldsymbol{x}_q\leq \epsilon$ and $\boldsymbol{J}^T_{m} = \boldsymbol{J}_{-m}$ then $\boldsymbol{x}_q^T \boldsymbol{J}^T_{m}\boldsymbol{x}_q\leq \epsilon$ also holds and from the first constraint $\boldsymbol{x}_q^T \boldsymbol{I}_{m}\boldsymbol{x}_q\leq p$). Now we can follow the steps of the algorithm below to design the training sequence.
\begin{algorithm}
\label{algo}
    \caption{CYCLIC OPTIMIZATION ALGORITHM FOR INTEGRATED TRAINING SEQUENCE DESIGN}
    \textbf{1}: Initialize $\boldsymbol{P}_{DL}$ and $\boldsymbol{P}_{UL}$ by assigning a matrix by random in the space $\Omega$.\\
    \textbf{2}: Update the minimizer $\boldsymbol{V}$ of Eq. (\ref{fvx}) using Eq. (\ref{minimizer}). \\ %using \ref{} 
    \textbf{3}: Update the current $\boldsymbol{X}$ and $\boldsymbol{Y}$ by solving cyclic optimization problem (\ref{XConv2}) and (\ref{YConv}) until convergence or $\mu$ times.\\
    \textbf{4}: Repeat the second and the third steps until the point that a stop criteria is fulfilled, e.g. $\left|MSE^{(m + 1)} - MSE^{(m)}\right|<\eta$ for some given $\eta > 0$, where $m$ means the external loop cycle.
	\addcontentsline{lot}{table}{\textbf{ALGORITHM I:} $\quad$ CYCLIC OPTIMIZATION ALGORITHM FOR INTEGRATED TRAINING SEQUENCE DESIGN}
\end{algorithm}

\section{Simulation Results}

The performance of the communication mode is evaluted in this section through simulations with respect to the channel MSE metric. Afterwards, for the radar mode we illustrate the cross and auto-correlation between two training sequences.

\subsection{Simulation Settings}

We used the exponential model to generate covariance matrices. This model is particularly appropriate whenever a control over correlation is required. We let $[\boldsymbol{C}]_{k,l} = \rho^{l-k}$ for $k \leq l$ for a covariance matrix $\boldsymbol{C}$, and $[\boldsymbol{C}]_{k,l} = [\boldsymbol{C}]_{l,k}^*$ for $l < k$, when $|\rho| < 1$ denotes the correlation coefficient. Furthermore, we assume that both the noise matrix $\mathbf{M}$ and the channel matrix $\mathbf{R}$  are following the Kronecker model; i.e for covariance matrix $\mathbf{R}$ defined as $\mathbf{R} = (\mathbf{R}^T_{\mathbf{T}} \otimes \mathbf{R_R})$ we suppose $\rho_{rt} = 0.91 e^{- j \theta_{rt} }$ and $\rho_{rr} = 0.6 e^{- j \theta_{rr}}$ to construct $\mathbf{R_T}$ and $\mathbf{R_R}$ (respectively at the transmit side and at the receive side) using exponential model. Also, for covariance matrix of noise $\mathbf{M}$ defined as $\mathbf{M} = (\mathbf{M}^T_{\mathbf{T}} \otimes \mathbf{M_R})$ where $\mathbf{M_R} = \mathbf{R_R}$, we let $\rho_{mt} = 0.8 e^{- j \theta_{mt}}$ to construct $\mathbf{M_T}$ at the transmitter side.The phase arguments $(\theta_{rt},\theta_{rr},\theta_{mt} )$ are chosen random with the values $(0.83\pi, 0.42 \pi, 0.53 \pi)$. 

Also $\mathbf{R}$ and $\mathbf{M}$ are normalized in a way that $tr\{\mathbf{R}\} = 1$ and $tr\{\mathbf{M}\} = 1$, also we define the SNR as 
$
\textrm{SNR} \triangleq \gamma
$, 
 and $\gamma=\| \boldsymbol{P} \|_F^2$ shows the \textit{total energy training phase}. 
 %Then for a given $\gamma$, different values of $\kappa$ realize different SNR values. 
 We consider $\gamma =  B n_T$, and set the stop criteria threshold of the loops for iterations in Algorithm I as  $\eta = 10^{-6}$.

\subsection{Channel MSE Metric} %}{Channel Estimate MSE and Design Criteria}

\begin{figure}
	\begin{center}
		\includegraphics[width=9.5cm]{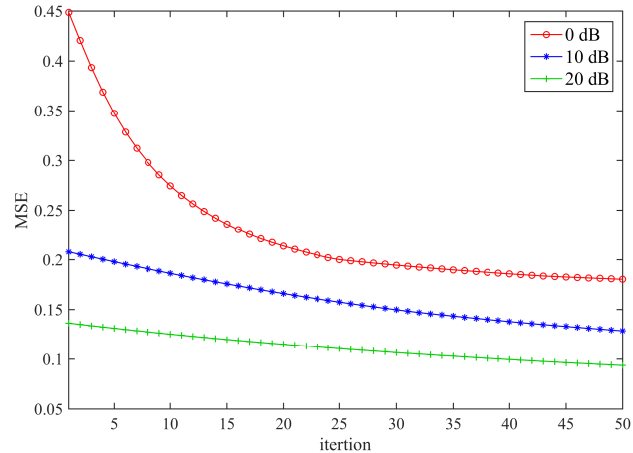}
		\caption{Comparison between different schemes with MIMO channel ($4 \times 4$) with $8$ training symbols per antenna vs MSE as the performance metric. To show the enhancement in the performance metric,  MSE values are shown at each iterations.} \label{fig:Fig1}
	\end{center}
\end{figure}

We show the performance of the suggested approach for communication purposes using MSE as the figure of merit.  The channel considered in this chapter is a $4 \times 4$ MIMO channel and the number of training data symbols per antenna $B=8$. The results are appeared in Fig.~\ref{fig:Fig1}. For each power, we have utilized the proposed strategy $ 50 $ times ,utilizing different settings, and have announced the average of the acquired MSE values. It can be seen from Fig.~\ref{fig:Fig1} that the proposed algorithm performs better in every iteration until the point when it focalizes to the ideal MSE.

\subsection{Radar Training Sequence Specifications}

\begin{figure}
	\begin{center}
		\includegraphics[width=9cm]{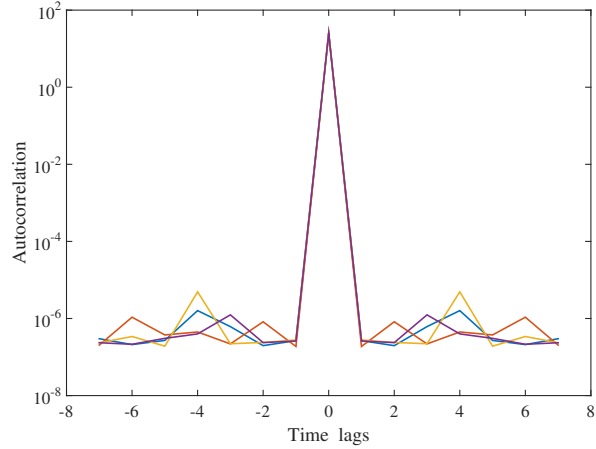}
		\caption{Autocorrelation of radar training sequence ($\boldsymbol{x}_l^T\boldsymbol{J}_i\boldsymbol{x}_l$) where $1\leq l\leq 8$, and each $l$ denotes a transmit antenna so we have totally 4 autocorrelation plots in this figure, and also  $-8\leq i\leq 8$ denote time lags} \label{fig:Fig2}
	\end{center}
\end{figure}

To ensure radar part of the system performs properly, our training sequence for radar part should have very small auto-correlation for at least a range of time lags so that this training sequence have an impulse like shape. In Fig. \ref{fig:Fig2} this auto-correlation is shown. For each lag, auto-correlation level is shown in dB. Fig. \ref{fig:Fig2} shows that auto-correlation levels for time lags 2-8 are almost zero compared to autocorrelation for the first lag. Which gives us the impulse-like correlation for the training sequence contributing in sensing mode.

\subsection{Correlation of training sequences}

\begin{figure}
	\begin{center}
		\includegraphics[width=9cm]{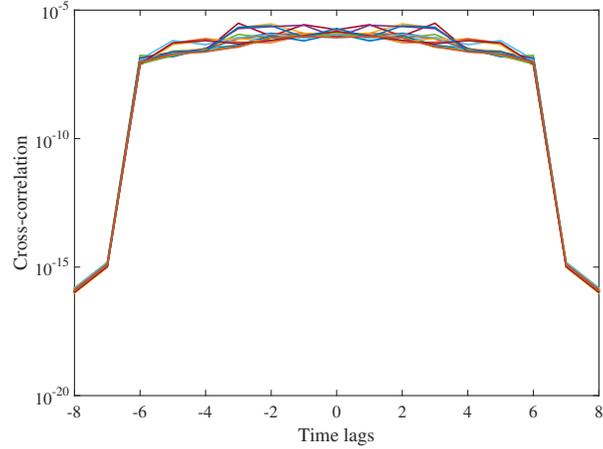}
		\caption{Cross-correlation of radar and communication training sequences ($\boldsymbol{x}_q^T\boldsymbol{J}_i\boldsymbol{y}_l$) where $1\leq q\leq 8$, and each $q$ denotes different transmit antenna, $1\leq l\leq 8$, and each $l$ denotes different receive antenna so we have totally 16 cross-correlation plots in this figure, and also  $-8\leq i\leq 8$ denote time lags} \label{fig:Fig3}
	\end{center}
\end{figure}

The key factor for our system to distinguish between radar signal and communication signal is that two training sequences should be uncorrelated with each other for a number of time lags. Fig. \ref{fig:Fig3} shows cross-correlation between two training sequences for our simulations in dB. As it is obvious from simulations results, correlation between these two signals are really small so they can be assumed uncorrelated.

\section{Conclusion}
The idea of designing training sequences for a communication system to be able to operate also in an integrated radar mode has been proposed and the protocol and limitations has been explained. We evaluate the channel MSE for communication system and show that training sequences are uncorrelated, and also one of the sequences have impulse-like correlation (suitable for radar sensing). The proposed system can perform as a radar and communication system. Considering the communication operation as the primary also
lays the ground for making the radar systems ubiquitous.

% Chapter 5

\chapter{Conclusion} % Main chapter title

\label{Chapter5} % For referencing the chapter elsewhere, use \ref{Chapter5} 

\section{Review of The Research}

With growing utilization of wireless communication systems, and also having limited spectral resources, using these resources more efficiently has become absolutely critical. That is the reason that this research has focussed on efficient ways of using spectral resources in wireless communication. This thesis proposed two main ideas to increase spectral efficiency. First, to change traditional optimization metrics, which was channel ergodic capacity, to more accurate maximal achievable rate approximations and use this metric to optimize the training data size for a wireless communication system. Using proposed accurate metric for optimization purposes results in more realistic optimization and of course more spectral efficiency, especially for short packet communication and ultra reliable applications. Second, to utilize training data to design a multifunction communication and sensing system in an integrated platform. This allows communication and sensing systems to share the same spectrum and in result achieve better spectral efficiency. Main findings of this research can be summarized in two main categories:

\subsection{Optimizing Training Data Size Using Maximal Achievable Rate Approximation}

This was our first solution to achieve better spectral efficiency which was discussed in Chapter \ref{Chapter3} where the idea of optimizing training data size using ergodic capacity (previously used method) was explained in detail. Then the proposed optimization metric which is maximal achievable rate approximation has been introduced. Finally, the idea of using this metric to optimize the training data size has been introduced. Furthermore, it was shown through simulation that our proposed method achieves higher performance especially when the packet size is small (short packet communication) and also when low probability of error is required (ultra reliable system). 

\subsection{Utilizing Training Data to Design a Multifunction Communication and Sensing System}

The second solution that has been provided in this thesis is to utilize the training data to design an integrated communication and sensing system. The details behind the proposed idea has been discussed in Chapter \ref{Chapter4}. The proposed integrated scheme has been introduced in Chapter \ref{Chapter4} in detail. We proposed that to integrate sensing abilities into an already built communication system, we can design a pair of training data sequences, one for the base-station, and one for the user. Then the design procedure for training data sequences is provided. It is shown through simulations that the designed training data sequences are perfectly uncorrelated for a range of time lags. Also simulation results shows that the training data sequence which is designed for sensing purposes has an impulse-like autocorrelation. This allows base-station to distinguish between them and in result it can utilize those sequences for communication and sensing proposes simultaneously.

\section{Implications of Findings}

Our research has specific aspects, in which its performance is superior to the prior works. Here we explain those aspects and also the real world applications of those aspects. 

\subsection{Better Performance in Short Packet Ultra Reliable Communication}

The idea of using maximal achievable rate approximation as the optimization metric to optimize training data size has shown better performance especially in the case of having a short packet (small $n$). This means the difference between our method of optimizing in compare with traditional methods is a considerable amount when the packet size is short. Also, simulation results in Chapter \ref{Chapter3} shows that this difference is considerable when low probability of error is required (small $\epsilon$). This situation happens in Wireless Sensor Networks (WSN), Internet of Things, Smart Homes, etc. Where sensors or “things” are connected together, they usually transmit small packets through communication network, but high certainty of securely receiving the packet is required. In such applications, our method works with considerably better performance.

\subsection{Ubiquitous Spectrum Sharing Between Communication and Sensing}

The proposed platform in Chapter \ref{Chapter4} to integrate sensing functionality into an already existing communication system has shown the ability for communication and sensing system to coexist in the same spectrum. Since the training sequences are designed to be uncorrelated to each other, they can be detected even in the same platform and also the same spectrum. Also, our proposed integrated system has more readily applications than previous attempts of integrated platforms, because of the fact that it is designed to add sensing ability to a communication system. Since nowadays communication systems are more ubiquitous than radar platforms, our proposed system can be used in a wide variety of applications. For instance, Mobile Network base-stations with adjustable beam-forming can use such sensing ability to shape the main lobe beam of their antenna array to enhance their performance.

\section{Limitaitons of Study}

The research presented in this thesis, although superior in some aspects, has some limitations compared with other works. Here the limitations of our study will be mentioned and the situations in which our proposed ideas should not be used will be discussed.

\subsection{Asymptotically Similar Performance for Long Packets}

The training data size is optimized using approximate maximal achievable rate as the optimization metric in Chapter \ref{Chapter3}. As it can be seen in simulation results in Fig. \ref{Ropt_cont}, the difference of performance between the proposed approach and the previous ones asymptotically goes to zero for long packet sizes ($n \rightarrow \infty$). The same phenomena occurs when higher probability of error is required ($\epsilon \rightarrow 1$). This suggest that our approach does not offer any advantage for long packet sizes or high error probabilities. Therefore, for applications with long packet size or high probability of error, using our proposed approach is not justifiable because it does not offer any advantage in those cases.

\subsection{Increased Computational Complexity}

It is common that in an attempt to enhance efficiency and accuracy, one can increase the computational complexity of a problem and this research is no exception. Here we discuss computational complexities of Chapters \ref{Chapter3} and \ref{Chapter4}.

\subsubsection{Complexity of Proposed Optimization Metric}

In Chapter \ref{Chapter3}, we replaced traditionally used optimization metric, ergodic capacity, with approximate maximal achievable rate expression. Just through looking at Eq. (\ref{32}), it can be seen that to optimize using such metric is more complex than optimizing the ergodic capacity. Therefore, our proposed method achieves more performance and rate with the cost of having increased complexity. This increased complexity could cause increased costs of implementation; however, with ever increasing processors speed, the issue of increased computational complexity is of lower importance nowadays.

\subsubsection{Complexity of Integrated Communication and Sensing System}

The idea of designing two distinct training data sequence is discussed in Chapter \ref{Chapter4}. To this end, in Eq. \ref{OP}, a whole new constrained optimization problem was introduced in every cycle. This increases computational complexity compared with the case when the system was only responsible for communication purposes. Thus, it can be seen again, that the advantage of adding sensing abilities to a communication system is gained with the cost of increasing computational complexity.

\section{Future Work}

In last section, the limitations of our study was discussed. Here we introduce some solutions in attempt to eliminate or minimize mentioned limitations.

\subsection{Analytical Analysis of Optimal Training Data Size}

In an attempt to decrease computational complexity of optimization with respect to maximal achievable rate expression, the most effective solution is to analytically solve the optimization problem. Although this specific optimization problem is almost impossible to solve analytically, using Taylor expansion and some other tight approximations, it can be solved analytically. Another tool that can be used to this end is to analyze extreme cases (low and high SNR regimes). The mentioned methods are being done in our current research. We used Taylor expansion to simplify the optimization problem to make it easier to be solved analytically and analyzed it in low and high SNR regimes. The results of this research is expected to be submitted in a foreseeable future.

\subsection{Complexity Trade-off for Integrated Communication and Sensing System}

As mentioned in previous section, adding sensing abilities to an already built communication system costs us increasing complexity. One way to deal with this problem is to have control over optimizing algorithm iterations. The integrated device does not need its sensing abilities to be accurate at all times. However, there are certain times when the sensing output is required to be accurate. The optimization algorithm can be designed in two different modes, one where the quality of communication system is the first priority, therefore, the optimization algorithm can use less iterations to increase speed with cost of loosing sensing accuracy. Other one where sensing is the first priority at which the algorithm can run more iterations to increase sensing accuracy with the cost of loosing speed.

%\subsection{Utilizing Proposed Ideas in Practical Applications}

%\myPart{III}{Controllers}{4cm}

%\myPart{IV}{Conclusion}{4cm}

\appendices
\newpage
\appendix

\chapter{Proof of Eq. (3.7) } \label{Appendix:doppler}

The user transmit $n_t$ training symbols known to the mobile and base-station, enabling the base-station to estimate the channel gain.
The MMSE channel gain estimator can be derived as \cite{mmse3}
\begin{eqnarray}
\widehat{\textbf{h}} &=& \sqrt{\rho} (1 + \rho|\textbf{x}_t|^2)^{-1} \textbf{x}^*_t \textbf{y}_t \nonumber\\
&=& \dfrac{1}{\sqrt{\rho}} \left( \dfrac{1}{\rho} + |\textbf{x}_t|^2 \right)^{-1} \textbf{x}^*_t \textbf{y}_t
\end{eqnarray}
where  $\textbf{x}_{t}$, $\textbf{y}_{t}$ are input and output training symbol vectors. Note that 
$\textbf{y}_t = \sqrt{\rho} \textbf{x}_{t} \textbf{h} + \textbf{w}_t + \sqrt{\rho} \textbf{x}_t\Delta\textbf{h}_t,$ where $\textbf{y}_t = [y(1), y(2), \dots, y(\alpha n)]$, $\textbf{x}_t = [x(1), x(2), \dots, x(\alpha n)]$, $\textbf{w}_t = [w(1), w(2), \dots, w(\alpha n)]$ are output, input and noise vectors respectively. Also $\Delta\textbf{h}_t = [\Delta h(1), \Delta h(2), \dots, \Delta h(\alpha n)]$ is the channel mismatch due to the temporal variation of the channel.  
Since $|\textbf{x}_t|^2 = \alpha n$, we get
\begin{eqnarray}
\widehat{\textbf{h}} &=& \sqrt{\dfrac{1}{\rho}} \left(\dfrac{\rho}{1 + \alpha n \rho} \right) \left[\sqrt{\rho} \alpha n \textbf{h} + \right. \nonumber\\
&& \left. \textbf{w} \textbf{x}^*_t + \sqrt{\rho} \Delta\textbf{h}_t  \textbf{x}^*_t\textbf{x}_t \right] \nonumber\\
&=& \left(\dfrac{\alpha n \rho}{1 + \alpha n \rho} \right) \textbf{h} + \sqrt{\dfrac{1}{\rho}} \left( \dfrac{\rho}{1 + \alpha n \rho} \right) \textbf{w}\textbf{x}^*_t \nonumber\\
&& + \left(\dfrac{\alpha n \rho}{1 + \alpha n \rho} \right) \Delta\textbf{h}_t.
\end{eqnarray}
Hence the channel estimation error  $\widetilde{\textbf{h}} = \textbf{h} - \widehat{\textbf{h}} $ can be derived as: 
\begin{eqnarray}
\widetilde{\textbf{h}} &=& \textbf{h} - \widehat{\textbf{h}} \nonumber\\
&=& \dfrac{1}{1+\alpha n \rho} \textbf{h} - \sqrt{\dfrac{1}{\rho}} \left( \dfrac{\rho}{1 + \alpha n \rho} \right) \textbf{w}_t\textbf{x}^*_t \nonumber\\
&& - \left(\dfrac{\alpha n \rho}{1 + \alpha n \rho} \right) \Delta\textbf{h}_t
\end{eqnarray}
Finally, we derive the mean square error of the MMSE channel estimation error as 
\begin{eqnarray}
\sigma^2_{\widetilde{\textbf{h}}} = \dfrac{1}{1+\alpha n \rho} + \sigma^2_{{Doppler}}
\end{eqnarray}
where we derived $\sigma^2_{{Doppler}}$  using the mathematical derivation proposed in \cite{deltah} for maximum likelihood (ML) channel estimator in Rayleigh fading channel
\begin{eqnarray}
\sigma^2_{Doppler} &=& 2\left(\dfrac{\pi\alpha n\rho f_D}{1 + \alpha n \rho}\right)^2\left( n - \dfrac{\alpha n}{2} \right)^2,
\end{eqnarray}
where $f_D$ is the Doppler frequency normalized to the symbol rate ($R_{symbol} = 1/T_s$) given by $\dfrac{T_s v f_c}{c}$, $T_s$ is the symbol period, $v$ is the mobile velocity, $f_c$ is carrier frequency and $c$ is the speed of electromagnetic wave.
%\section*{Acknowledgment}
%Funding for this research was provided by NSF-CCF (Award Number 1320419).

%\chapter{Copyright Permissions}  \label{Apndx:CopyRight}
%In this appendix, we present the copyright permissions for the articles, whose contents were used in this thesis. The list of the articles include Humanoids'14 \cite{Humanoids}, WHC'15 \cite{WHC_MinJerkCoop} and RO-MAN'16 \cite{myROMAN} conference papers following by an article in IEEE Transactions on Robotics \cite{myHRI_TRO} and a book chapter in Springer \cite{myBookCh}. The permissions follow in the same order.
%
%\RemoveIFiThenticate{
%%\includepdf[pages=-, fitpaper=true]{./Appendices/Humanoids_Copyright.pdf}
%%\includepdf[pages=-, fitpaper=true]{./Appendices/WHC_Copyright.pdf}
%%\includepdf[pages=-, fitpaper=true]{./Appendices/ROMAN_Copyright.pdf}
%\includepdf[pages=-, fitpaper=true]{./Appendices/TRO_Copyright.pdf}
%%\includepdf[pages=-, fitpaper=true]{./Appendices/Sprngr_Copyright.pdf}
%}

\chapter{Author's Biography}  \label{Apndx:Biography}

Mohammadreza Mousaei was born and raised in Tehran, Iran. He received his B.Sc. in Electrical Engineering from Shahid Beheshti University. Thereafter, He worked at Sharif University of Technology's Computer Vision Laboratory and Amirkabir University of Technology's  Robotics Laboratory as a research assistant before he moved to U.S. and attend University of Illinois at Chicago to start his Masters. He is currently a M.Sc. candidate under supervision of Dr. Besma Smida.

His research interests span over areas of Robotics, Computer Vison as well as Information Theory and Wireless Communication. 

\nocite{}
\bibformb
\bibliography{Dissertation}

\newpage
\vita
\RemoveIFiThenticate{
\begin{singlespace}

\centerline{\huge Mohammadreza Mousaei}
\vspace{11pt}

%\RemoveIFiThenticate{
%\includepdf[pages=-, fitpaper=true]{./CV/cv.pdf}
%}
%%
\begin{list}
{}
{\setlength
   {\labelwidth}{1in}
    \setlength{\leftmargin}{1.5in}
    \setlength{\labelsep}{.5in}
    \setlength{\rightmargin}{0in}}
\item[Education\hfill] {
M.Sc. Electrical and Computer Engineering \hfill 2015 -- 2017\\
University of Illinois at Chicago, Chicago, IL

B.S. Electrical Engineering \hfill 2010 -- 2015\\
Shahid Beheshti University, Tehran, Iran \\
}

\item[Publications\hfill] 
{\textbf{M. Mousaei} and B. Smida, ``Optimizing Pilot Overhead for Ultra-Reliable Short-Packet Transmission'',  International Conference on Communication, IEEE, 2017.
\\

\textbf{M. Mousaei}, M. Soltanalian, B. Smida``ComSens: Exploiting Pilot Diversity for Pervasive Integration of Communication and Sensing in MIMO-TDD-Frameworks'', Military Conference on Communication, IEEE, 2017.
\\

A. Sheikhjafari, S. Gharghabi, K. Sartipi, E. Babaians, \textbf{M. Mousaei}, S. Shiry Ghidary, ``Amirkabir
University of Technology (AUT) @Home 2014 Team Description Paper'', Robocup 2014, Joao Pessoa, Brazil, 2014.
\\

A. Keipour, K. Sartipi, \textbf{M. Mousaei}, S. Mohammadzadeh, M. Jamzad, ``Team Description Paper for Sharif University
of Technology (SUT) Team'', AUTCUP Robotics competitions,  2013.
\\

%\textbf{Author} and Author, ``Paper Title'', Conference Name, Pages, City, Date.\\
}

\item[Awards\hfill] {
	
	\begin{itemize}
		\item 2nd Place, AUT-CUP Robotic International Competition, Artificial Intelligence league, 2014
		\item 2nd Place, ACM ICPC Qualiffication Programing Contest, 2012
		\item 19th Place, ACM ICPC West Asia Regional Programing Contest , 2012
		\item 5th Place, ACM ICPC Qualiffication Programing Contest, 2011
		\item Ranked Top 1\%, Mathematics and Physics among more than 178,000 students in Iranian nationwide
		university entrance examination (Konkoor).
		\item Accepted in Iranian National Olympiad Competition in Mathematics, 2008
		\item Accepted in Iranian National Olympiad Competition in Computer Science, 2007
	\end{itemize}
	
	}
%
%\item[Presentations\hfill] 
%{Invited Talks at ....\\
%Conference Presentations at ....\\
%Poster Presentations at ....\\
%}
%
%\item[Memberships\hfill] {include a list .... \\}
%
%\item[Services\hfill] 
%{Journal Article Referee at ....\\
%Conference Paper Referee at ....\\
%Graduate Employee's Organization (GEO) Steward at ...\\
%Student Volunteer at ....\\
%}
%
%\item[Experience\hfill] 
%{Visiting Researcher at ....\\
%Research Assistant at ....\\
%Summer Intern at ....\\
%Mentor for Undergraduate Students at ....\\
%Adjunct Lecturer at ....\\
%Teaching Assistant at ....\\
%}

\end{list}

\end{singlespace}
}

\end{document}